\documentclass[12pt,draftcls,onecolumn]{IEEEtran}
\usepackage{amsmath,amssymb,amsfonts,mathrsfs,bm}
\usepackage{amstext}
\usepackage{upgreek}
\usepackage{multicol}
\usepackage{graphicx}
\usepackage{paralist}
\usepackage{hyperref}
\usepackage[numbers,sort&compress]{natbib}
\usepackage{booktabs}
\usepackage{multirow}
\usepackage{subfigure}

\newtheorem{proposition}{Proposition}

\IEEEoverridecommandlockouts

\begin{document}

\title{Cache-enabled Device-to-Device Communications: Offloading Gain and Energy Cost}
\author{
\IEEEauthorblockN{{Binqiang Chen, Chenyang Yang and Andreas F. Molisch}}
\thanks{
	Binqiang Chen and Chenyang Yang are with the School of Electronics and Information Engineering, Beihang University, Beijing, China, Emails: chenbq@buaa.edu.cn, cyyang@buaa.edu.cn. Andreas F. Molisch is with the Ming Hsieh Department of Electrical Engineering, University of Southern California, Los Angeles, CA, USA. Email: molisch@usc.edu.
	}
%
}

\maketitle
\vspace{-8mm}\begin{abstract}
By caching files at users, content delivery traffic can be offloaded via device-to-device (D2D) links if a helper user is willing to transmit the cached file to the user who requests the file. In practice, the user device has limited battery capacity, and may terminate the D2D connection when its battery has little energy left. Thus, taking the battery consumption allowed by the helper users to support D2D into account introduces a reduction in the possible amount of offloading. In this paper, we investigate the relationship between offloading gain of the system and energy cost of each helper user. To this end, we introduce a user-centric protocol to control the energy cost for a helper user to transmit the file. Then, we optimize the proactive caching policy to maximize the offloading opportunity, and optimize the transmit power at each helper to maximize the offloading probability. Finally, we evaluate the overall amount of traffic offloaded to D2D links and evaluate the average energy consumption at each helper, with the optimized caching policy and transmit power. Simulations show that a significant amount of traffic can be offloaded even when the energy cost is  kept low.
\end{abstract}\vspace{-5mm}

\vspace{-2mm}\section{Introduction}
\label{s:1}
Device-to-device (D2D) communications boosts the throughput of cellular networks by offloading traffic \cite{doppler2009device,Andrews.D2D,zhang2013exploring,Andreev.JSAC}, and thus is a promising way to achieve the goal of 5th generation (5G) mobile networks.
Traditional D2D communication, which does not cache content locally, can only offload peer-to-peer (P2P) traffic from cellular networks if source and destination are in proximity at the time they wish to communicate,  such as gaming and relaying \cite{doppler2009device,Andrews.D2D,zhang2013exploring,Andreev.JSAC,Survey.D2D}. However, the lion's share of cellular traffic is video dissemination, a kind of client/server (C/S) services, which will generate more than $2/3$ of mobile data traffic by 2019 \cite{CISCO}.

Motivated by the fact that a large amount of content requests are asynchronous but redundant, i.e., the same content is requested at different times, caching at the wireless edge has become a trend for content delivery, which improves the throughput and energy efficiency of the network and the quality of 
experience (QoE) of the users \cite{wang2014cache,Procach14,Ali13,Dong,Higgins12,LHui14,Chen15}.


%

Recent work \cite{Mo.Mag13,Golrezaei.TWC} has shown that caching at the user devices enables offloading also of C/S traffic, in particular video, to D2D connections. Without caching at the devices, the users
need to fetch their requested video via base station (BS) from a remote server. By pre-downloading popular files to
users during the off-peak time, say at night, the file requested by a user can be
transmitted via D2D links by other users in proximity that have cached the file. Such a proactive caching policy largely alleviates the burden to the BSs during the peak time, yielding high offloading gain \cite{Mo.Mag13,Golrezaei.TWC,JJJ.JSAC,JMY.JSAC,Plac.D2D}. To improve the performance of cache-enabled D2D communications, proactive caching policies were optimized in \cite{JMY.JSAC,Plac.D2D}, and a distributed reactive caching mechanism was designed in \cite{JJJ.JSAC}.

When D2D communications are used for supporting P2P services, the users acting as transmitters are by definition willing to send messages to the destination users. However, offloading  content delivery traffic by
cache-enabled D2D communications needs the help of other users who are not obligated
to help. Due to the limited battery capacity, a natural question from a helper user in
such a network is: ``why should I spend energy of \emph{my} battery to provide \emph{you} with faster video download?
\cite{Mo.Mag13}"  This makes the energy consumption of a helper user a big concern in cache-enabled D2D communications. In practice, a helper user may only be willing to use a fraction of its battery for transmitting files to other users, if properly rewarded by the operator.
It is thus important to quantify the offloading gain when the helper
users' allowed battery consumption is taken into account, and to evaluate the average energy consumed by a
helper user to deliver the files to others.

In previous research efforts for cache-enabled D2D communications \cite{Mo.Mag13,Golrezaei.TWC,JMY.JSAC,JJJ.JSAC,Plac.D2D}, the energy of the battery is implicitly assumed infinite and the energy costs at helper users are never considered. Consequently, (i): maximal transmit power is used by all D2D transmitters to deliver the files, and (ii): once a D2D link is established, the file is assumed to be able to be delivered completely without considering whether there is still energy in the battery or whether a helper is willing to contribute more energy.

In this paper, we quantify the offloading gain of a cache-enabled D2D communication system by taking maximal permissible battery consumption into account, and evaluate the energy cost for a user to transmit the file. 
With the allowed battery consumption, a helper user may only transmit part of a file to the user requesting the file. To control the energy spent by the helper user for transmitting a file, we consider a user-centric caching and transmission strategy, where only the users within a \emph{collaboration
	distance} $r_c$ of the requesting user can serve as helpers. When the collaboration distance is large, the
probability that the users can fetch their desired contents via D2D links is high, and thus more traffic can be
offloaded. However, since the possible D2D link distance increases, the energy cost of a helper user also grows and then more files cannot be conveyed completely via D2D links.

Aimed to find the maximal offloading gain, we first introduce a user-centric probabilistic caching policy, where the users proactively cache files according to a $r_c$-dependent caching distribution. We optimize the policy to maximize the amount of traffic that can be possibly offloaded with
a given collaboration distance and the user demands statistics. In \cite{JMY.JSAC}, a \emph{cluster-centric} caching policy was proposed, which was optimized to maximize the same objective with given cluster size and demands statistics, but is not optimal under the user-centric framework. Then, we optimize the transmit power at each helper to
maximize the probability
that a requested file can be found in adjacent users and transmitted completely via a D2D link, considering two extreme cases in terms
of interference level. Finally, we quantify the total offloaded amount of traffic by taking complete and partial transmission into account, evaluate the average energy consumption for each D2D transmitter with optimized caching policy and transmit power, and characterize the relationship between  offloading gain and energy cost.

The contributions of this paper are summarized as follows:
\begin{itemize}
	\item  We analyze  the offloading gain when the user only allows partial energy in its battery to be consumed. To the best of the authors' knowledge, this is the first paper to characterize the offloading gain given limited battery consumption in cache-enabled D2D communications.
	\item We investigate the relationship between the offloading gain of the system and the energy costs of the helper user, and show the impact of the allowed battery consumption.
\end{itemize}

The rest of the paper is organized as follows. Section \ref{s:2} presents the system model. Section \ref{s:3} optimizes the caching policy. Section \ref{s:4} optimizes the transmit power, and evaluates the offloading gain and energy cost. Section \ref{s:5} shows simulations. Section \ref{s:6} concludes the paper.

\section{System Model}
\label{s:2}
Consider a cell where users' locations follow a
Poisson Point Process (PPP) with density $\lambda$.
Each single-antenna user has local cache to store files, and can act as a helper  to transmit but with only a fraction of its battery capacity.
For simplicity of notation, assume that each user only stores one file in its local cache as in \cite{Golrezaei.TWC,Plac.D2D}, though generalization to storage of multiple files is straightforward.

When a helper transmits a file in the local cache via D2D link to a user requesting the file, i.e., a D2D receiver (DR), the helper becomes a D2D transmitter (DT). To control the energy spent by a DT for transmitting to a DR, we introduce a \emph{user-centric} protocol. A DT will send a cached file to the DR only if their
distance is smaller than a given value $r_{c}$, called \emph{collaboration distance}. The users with distance $r$ less than $r_c$  are called \emph{adjacent users}. Assume that a fixed bandwidth is assigned to the D2D links to avoid the interference between D2D and cellular links \cite{Survey.D2D}, and all DTs transmit with same transmit power. The BS is aware of the  files cached at the users and coordinates the D2D communications.

\vspace{-0.2cm}
\subsection{Content Popularity and Caching Placement}
We consider a static content catalog consisting of $N_f$ files that all users in the cell may request, which are
indexed in descending order of popularity, i.e., the 1st file is the most popular file. Each file has size of $F$ bits, but the analysis can be easily extended to general cases by dividing each file into chunks of equal size. The probability that the $i$th file is requested follows a Zipf distribution
\begin{equation}
\label{equ.p_r}
\textstyle
p_r(i)=\frac{i^{-\beta}}{\sum_{k=1}^{N_f}k^{-\beta}},
\end{equation}
where $\sum_{i=1}^{N_f}p_r(i)=1$, and the parameter $\beta$ reflects how skewed the popularity distribution is, with large $\beta$ meaning that a few files are responsible for the majority of requests \cite{Zipf99}.

\begin{figure}[!htb]
	\centering
	\includegraphics[width=0.35\textwidth]{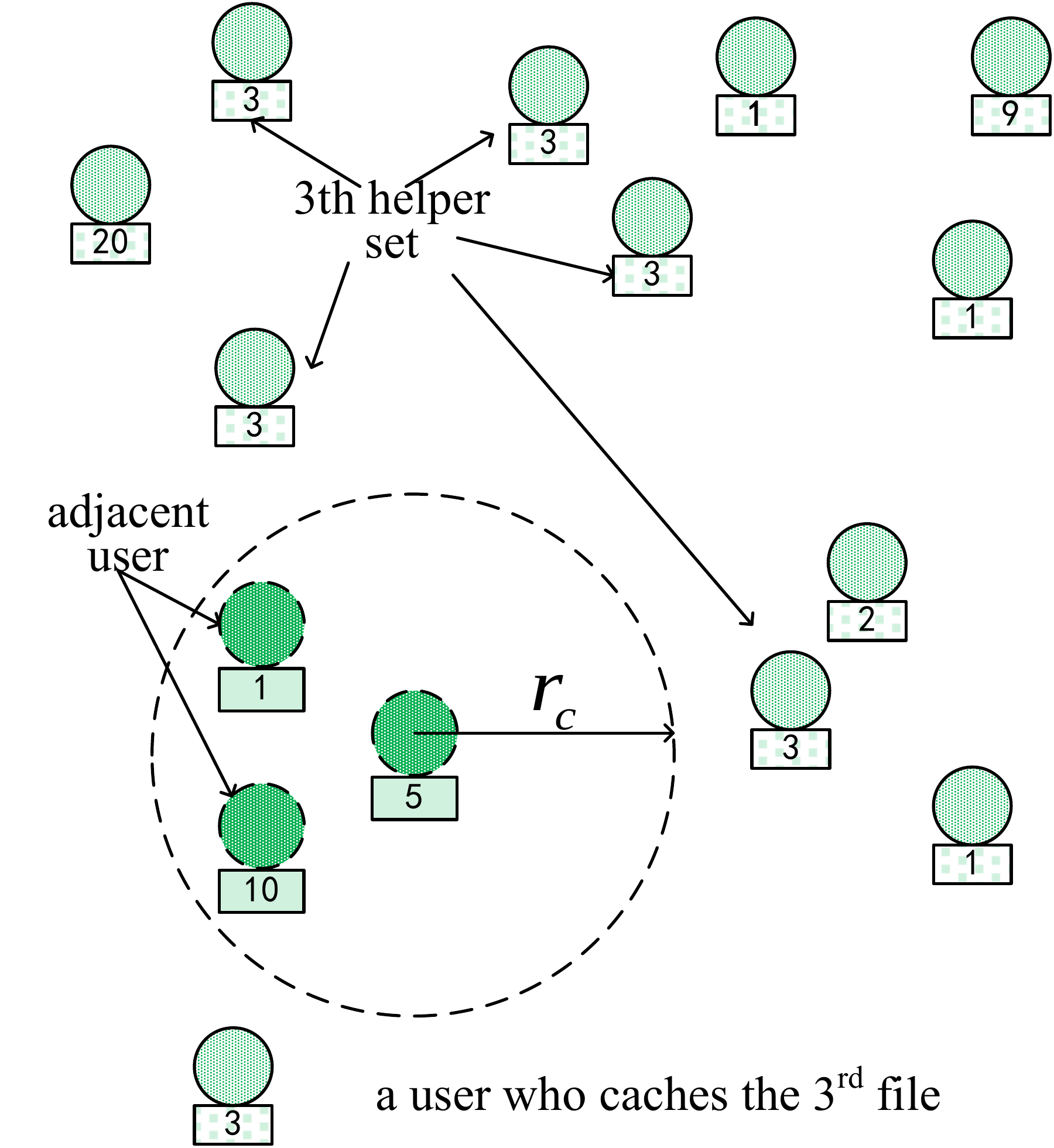}\\\vspace{-0.5cm}
	\caption{Illustration for a user-centric cache-enabled D2D network}\label{fig.2}
\end{figure}

Since deterministic caching policy designed for wired networks with fixed topology is not applicable for a wireless scenario with user locations that are unknown \emph{a priori}, we consider a probabilistic caching policy. Specifically, each user caches a file according to a $r_c$-dependent caching distribution, which is the
probability that the $i$th file is  cached at users, $i=1,\cdots,N_f$. All users in the cell that have cached with the $i$th
file constitute a user set, called the $i$th \emph{helper set}, as shown in Fig. \ref{fig.2}.

\subsection{User Allowed Battery Consumption and Content Delivery}

 The content popularity usually changes at a much slower speed than the traffic variation of cellular networks (e.g., one week for movies \cite{Mo.Mag13}), which is often regarded as invariant over a period. Consequently, the files can be proactively downloaded
by the BS during the off-peak time, without the need to be updated frequently. The energy consumed at users during content placement is negligible since users will usually be connected to the AC power during the download time (say at night).

Assume that each user requests one file from the catalog independently. If a user can find its requested file in the local caches of its \emph{adjacent users}, a D2D link is established
between the user and its nearest adjacent user cached with the file to convey the file. Assume that each user device has the same battery capacity of $Q$ (mAh), and only a fraction $\rho$ of each DT's battery capacity can be consumed for transmitting a file to the DR. Denote the operating voltage of the user device as $V_0$. When a  DT  has consumed $\rho QV_0$ energy to transmit a file to a DR, the DT interrupts the D2D link, and the DR has to receive the remaining data of the file from the BS.\footnote{$\rho$ can reflect the user incentive in terms of battery consumption to serve as a helper. We assume that all users are initially with full battery and hence each user allows to employ the same amount of energy to help others. In practice, the user devices may have different battery capacities. Moreover, a helper may be requested more than once over several hours before  recharging its battery, especially when the file is very popular.  When a DT serves the second request, the remaining energy in its battery may be less than $\rho QV_0$. For analysis simplicity, we assume that one user only sends one request, and hence each DT is only requested once. The impact that a DT serves multiple requests will be shown via simulation later.}  In fact, another helper in the adjacent of the DT can take over the transmission. We do not consider the hand over among DTs due to the following reason. The distances between the DR and other not-busy helpers are always longer than the distance between the DR and its first-established DT, and hence the corresponding channel conditions are worse in high probability (e.g., when $r_c$ = 100 m and $\beta = 1$, this probability is 97 \%). As a result, the handover will introduce higher energy cost for other DTs and more signaling overhead for the BS to coordinate. Therefore, two cases may occur for the established D2D links depending on their channel conditions.
\begin{itemize}
	\item { \bf Complete transmission:} A DR can receive a complete file via D2D link, which is called a \emph{satisfied DR}.
	
	\item {\bf Partial transmission:} A DR only receives a fraction of the file from a DT, which needs to access to the BS to fetch the remaining file.
\end{itemize}

If a user cannot find its requested file in the local caches of its \emph{adjacent users},  the user fetches the file from the BS.
If a user can find the desired file in its own local cache, such a self-serve can offload traffic without establishing D2D link. Since we focus on the energy cost of a DT in cache-enabled D2D communications, we ignore self-serve (also called self-offloading in literature) in the forthcoming analysis (similar to \cite{Golrezaei.TWC,JMY.JSAC,JJJ.JSAC,Plac.D2D}), but we will evaluate its impact via simulations in section V.

We consider two metrics regarding offloading by the cache-enabled D2D communications.
\begin{itemize}

	\item { \bf Offloading probability:} This is the probability that a DR enjoys complete transmission via D2D links, which reflects the percentage of the satisfied users.
	\item { \bf Offloading ratio:} This is the ratio of the amount of data offloaded by both complete and partial transmission via D2D links to the total amount of data in the cell, which reflects the offloading gain of the system.
\end{itemize}
To focus on the energy cost issue, we assume that the distance between DT and DR remains fixed during transmission (again following  most of previous works \cite{Mo.Mag13,Golrezaei.TWC,JMY.JSAC,JJJ.JSAC,Plac.D2D} ), although user mobility is one of the key factors that affects the offloading gain of cache-enabled D2D communications.


\section{Optimal Caching policy}
\label{s:3}
To optimize the probabilistic caching policy with known user demand statistics, we need to find the optimal caching distribution. Because the contents are proactively placed at users before they initiate requests, we optimize the caching distribution to maximize \emph{offloading opportunity} as in \cite{Mo.Mag13,Golrezaei.TWC,JMY.JSAC,JJJ.JSAC}, defined as the probability that the desired file of a user can be found in {adjacent users}. Such an opportunity reflects how much traffic can be \emph{possibly} offloaded by D2D communications  for a given collaboration distance under the assumption of infinite battery capacity. 

Denote the probability that the $i$th file is cached at a user as $p_c(i)$. Then, the set $\{p_c(i)\}=[p_c(1),p_c(2),...,p_c(N_f)]$ constitutes the caching distribution. The locations of the users who belong
to  the $i$th \emph{helper set} follow a PPP with density $\lambda_i = \lambda p_c(i)$ according to the thinning property of PPP \cite{SKM.PPP}.
Thus, the probability that a user requesting the $i$th file can find its desired file in the cache of any user within
the collaboration distance $r_{c}$ is $p_f(i) =1-e ^{-\lambda_i \pi r_{c}^2}$. Then, the offloading opportunity with given caching distribution and $r_c$ can be derived as
\begin{equation}
 \label{equ.p_o}\textstyle
{p}_o=\sum_{i=1}^{N_f}p_r(i)p_f(i)=\sum_{i=1}^{N_f}p_r(i)(1-e ^{-\lambda p_c(i) \pi r_{c}^2}).
\end{equation}

The optimal caching distribution that maximizes the offloading opportunity can be found from the following problem
\begin{equation}
\label{equ.opt1}
\begin{aligned}
\max_{p_c(i)} \,\, & \textstyle{p}_o\\
 s.t.\quad
&\textstyle\sum_{i=1}^{N_f} p_c(i)=1, \quad p_c(i) \geq 0, \quad i=1,\cdots, N_f.
\end{aligned}
\end{equation}
Because the objective function is the sum of $N_f$ exponential functions and the constraints are  linear, this problem is
convex  \cite{SL.OPT}.
It is not hard to show from its Karush-Kuhn-Tucker (KKT) conditions that the optimal caching distribution should satisfy the
following conditions
\begin{equation}
\label{equ.Lag_3}
\begin{split}
\textstyle
p^*_c(i) = \left[\frac{1}{\lambda \pi r_{c}^2}\ln(p_r(i))-\frac{1}{\lambda \pi r_{c}^2}\ln(\frac{-\mu}{\pi \lambda
r_{c}^2})\right]^+,
\end{split}
\end{equation}
where $1 \leq i \leq N_f$, $\sum_{i=1}^{N_f} p^*_c(i) =1$, $p_r(i)$ is the Zipf distribution in \eqref{equ.p_r}, and $[x]^+ = \max(x,0)$.

\begin{proposition} \label{p:1}
If $\frac{(N_f)^{N_f}}{N_f!} < e^{\frac{\lambda \pi r_{c}^2}{\beta}}$, then the optimal caching distribution is
\begin{equation} 
\label{equ.p:1.1}\textstyle
p^*_c(i) =   \frac{1}{N_f} \left(1+ \frac{\beta}{\lambda \pi r_{c}^2}\sum_{j=1}^{N_f}\ln(\frac{j}{i})\right).
\end{equation}
Otherwise, the optimal caching distribution is
\begin{eqnarray}
\label{equ.p:1.2}\textstyle
p^*_c(i)=
\begin{cases}
\frac{1}{i^*} \left(1 + \frac{\beta}{\lambda \pi r_{c}^2 }\sum_{j=1}^{i^*}\ln(\frac{j}{i})\right), &i \leq i^*, \\
0, &i^*<i \leq N_f,\\
\end{cases}
\end{eqnarray}
where $i^*$ is upper and lower bounded as $\frac{\lambda \pi r_{c}^2}{\beta} -1 \leq i^*
\leq \frac{\lambda \pi r_{c}^2}{\beta} + \ln(\sqrt{2\pi N_f}) + 1$.
\end{proposition}

\begin{IEEEproof}
See Appendix \ref{a:p1}.
\end{IEEEproof}

\vspace{2mm} The gap between the upper and lower bounds of $i^*$ in \eqref{equ.p:1.2} is $\ln(\sqrt{2\pi N_f})+2$, which is small. For example, when $N_f=1000$, the gap equals to $4.4$. This suggests that $i^*$ and hence the optimal caching distribution $p^*_c(i), i=1,\cdots, N_f$ can be obtained efficiently. $p^*_c(i)$
depends on the collaboration distance $r_c$, user density $\lambda$, as well as content statistics $N_f$ and $\beta$.

When $r_c \rightarrow \infty$, $\frac{(N_f)^{N_f}}{N_f!} < e^{\frac{\lambda \pi r_{c}^2}{\beta}}$ holds, and according to \eqref{equ.p:1.1} $p^*_c(i) = \frac{1}{N_f}$. In this case, the optimal caching distribution is a uniform distribution, i.e., each user can randomly choose a file to cache, because the number of adjacent users for any user trends to infinity.

By using the conditions below \eqref{equ.p:1.2} and setting $i^*=N_f$, it is not hard to show that when $r_c \leq \sqrt{\frac{(N_f+1)\beta}{\pi\lambda}} $, $\frac{(N_f)^{N_f}}{N_f!} \geq e^{\frac{\lambda \pi r_{c}^2}{\beta}}$. In this case, $p^*_c(i)$ is computed with \eqref{equ.p:1.2}, and the less popular files with indices larger than $i^*$ are never cached at the users. Because the number of adjacent users are limited when $r_c$ is small, only the files with high popularity are cached.
When $r_c \rightarrow 0$, $p_c^*(1)=1$ and $p_c^*(i)=0$, $ 1 <i \leq N_f$, i.e., only the most popular file is cached at each user.



\section{Offloading Gain and Average Energy Costs}
\label{s:4}
In this section, we investigate the offloading gain of the system and the energy cost at each DT. To this end, we first optimize the transmit power of each DT  to maximize the \emph{offloading probability}, which yields maximal user satisfaction rate and hence high offloading gain. Then, we evaluate the \emph{offloading ratio} and the average energy consumed at each DT to transmit a file via D2D links with the  optimized transmit power and optimized caching policy.

Considering that the interference among D2D links has large impact both on the offloading gain and the energy cost, for mathematical tractability we analyze two extreme cases in terms of interference level: full reuse and time division multi-access (TDMA). With full reuse, all DTs in a cell simultaneously transmit over the time and frequency resources are assigned for D2D communications without any interference coordination. With TDMA, only one DT in the whole cell transmits at a time, and the DTs are scheduled according to round robin (or random) scheduling with equal time slot duration.  While further improvements could be achieved through scheduling, it is known that optimal scheduling in D2D networks is NP-hard. On the other hand, cluster-based scheduling as in \cite{Golrezaei.TWC} is not aligned with the user-centric transmission strategy that forms the basis for our model.


\subsection{Case 1: Full Reuse}
\label{s:4.1}
Once a D2D link is established, the DT can transmit its cached file to the DR that requests the file.
In the full reuse case, each DR treats the interference among the D2D links as noise when
decoding the desired signal. The signal to interference plus noise ratio (SINR) at the DR requesting the $i$th file from its corresponding DT is
\begin{equation}
\label{equ.SINR_1}\textstyle
\gamma_1(i,r)  = \frac{P_thr^{-\alpha}}{\sum_{j \ne i} P_t h_j r_j^{-\alpha} +\sigma^2} = \frac{hr^{-\alpha}}{I_{i,r}+\sigma_0^2},
\end{equation}
where $P_t$ is the transmit power at each DT, $h$ is the channel power gain that follows an exponential distribution with unit mean for  Rayleigh fading, $r$ is the distance between the DT and the DR, $\alpha$ is the path loss exponent, $I_{i,r} = \sum_{j \ne i} h_j r_j^{-\alpha} $ is the total interference from other DTs  normalized by  $P_t$, $\sigma^2$ is the variance of white Gaussian noise, and $\sigma_0^2 = \sigma^2/P_t$.\footnote{Note that this model neglects shadowing and incorporating shadowing would lead to a change of the exponential channel gain distribution to an approximate lognormal distribution \cite{Mehtaetal}. We neglect shadowing, in line with most works in D2D literature.}  Then, the data rate is
$R_1(i,r)  =  W\log_2\left(1+\frac{hr^{-\alpha}}{I_{i,r}+\sigma_0^2}\right)$,
where $W$ is the bandwidth assigned to D2D links.

To evaluate the energy cost of each DT, we consider both circuit power and transmit power.
Then, the energy consumed  to transmit the $i$th file via a D2D link with distance $r$  is
\begin{equation} \label{equ.E_1}\textstyle
E_1(i,r) = \frac{F}{W\log_2\left(1+\frac{hr^{-\alpha}}{I_{i,r}+\sigma_0^2}\right)}\left(\frac{1}{\eta}P_t+P_c\right),
\end{equation}
where $\eta$ is the power amplifier efficiency and $P_c$ is the circuit power at the DT.

Because only a fraction $\rho$ of the battery capacity is permitted to be used at each DT to help a DR, a DT can transmit the $i$th file completely only if $E_1(i,r) \leq \rho V_0 Q$.

\subsubsection{Optimal Transmit Power}
Because the files not completely delivered via D2D links need to be fetched from the BS, which not only introduces extra signaling overhead but also may degrade the user experience, we optimize the transmit power at a DT to maximize the user satisfaction rate.
In other word, we maximize the offloading probability for a given collaboration distance $r_c$, which is the probability that a requested file can be found in adjacent users and transmitted completely via a D2D link.
\begin{proposition}\label{t:1}
The offloading probability in the full reuse case is
\begin{equation}
\label{equ.t:1.1}\textstyle
p_1(P_t,\rho) = \sum_{i=1}^{N_f} p_r(i) \int_{0}^{r_c} f_i(r) e^{- \phi_i(\Gamma_1,r)} dr,
\end{equation}
where $\Gamma_1=e^{\frac{F (P_t + \eta P_c)\ln2}{W\rho Q V_0 \eta }} -
1$, $f_i(r)=2\pi r \lambda_i e^{-\lambda_i \pi r^2}$ is the probability density function (pdf) of the D2D link distance, $\phi_i(x,y)\triangleq xy^\alpha \sigma_0^2 + \pi (\lambda_I\xi_1-\lambda_i^d\xi_2) y^2 x^{2/\alpha} $, $\lambda_I = \sum_{i=1}^{N_f} \lambda_i^d$ is the density of all DTs and
$\lambda_i^d = \lambda_i \left(1 - \left(1+\frac{\lambda p_r(i)}{3.5\lambda_i} \right)^{-3.5} \theta_i \right)$ is the density of DTs cached with the $i$th file,  $\theta_i = \frac{\Gamma(3.5,0)-\Gamma\left(3.5,\left(3.5\lambda_i+\lambda p_r(i)\right) \pi r_c^2\right)}{\Gamma(3.5,0)-\Gamma(3.5,3.5\lambda_i \pi r_c^2)}$,
$\Gamma(s,x) = \int_{x}^{\infty}t^{s-1}e^{-t}dt$ is the upper incomplete gamma function \cite{abramowitz1964}, $\xi_1 \triangleq \int_{0}^{+\infty} \frac{1}{1+t^{\alpha/2}}dt$, and $\xi_2 \triangleq \int_{0}^{x^{-\frac{2}{\alpha}}} \frac{1}{1+t^{\alpha/2}}dt$.
\end{proposition}

\begin{IEEEproof}
See Appendix \ref{a:t1}
\end{IEEEproof}\vspace{2mm}

The expression in \eqref{equ.t:1.1} depends on the values of $\lambda$, $r_c$, $\rho$ and $P_t$, but not on the user's location and channel.
To maximize the offloading probability for the cache-enabled D2D communications with given values of $\lambda$, $r_c$ and $\rho$, the transmit power at each DT can be optimized as
\begin{equation}
\label{equ.opt2}
\begin{aligned}
\max_{P_t} \,\, & \textstyle p_1(P_t,\rho)\\
s.t.\quad
& \textstyle 0< P_t \leq P_{\max},
\end{aligned}
\end{equation}
where $P_{\max}$ is the maximal transmit power of a DT.

Due to the complicated expression of $p_1(P_t,\rho)$, in general the optimal solution $P^*_t$ can only be found by using similar method as in \cite{Golrezaei.TWC}. When $r_c$ is small, all D2D links experience a line of sight (LOS) environment  \cite{JMY.JSAC}, i.e., $\alpha=2$. In such a special case, both closed-form expressions of $p_1(P_t,\rho)$ and $P_t^*$ can be obtained.

\begin{proposition}\label{sp:1} When $\alpha=2$, the offloading probability can be approximated as
	\begin{equation}
	\label{equ.sp:1.1}\textstyle
	p_1(P_t,\rho)\textstyle \approx \sum_{i=1}^{N_f}  \frac{p_r(i) \pi\lambda_i}{\varphi_i(P_t)}(1 - e^{- \varphi_i(P_t) r_c^2}),
	\end{equation}
which first increases and then decreases with $P_t$, where $\varphi_i(P_t)$ is defined in \eqref{equ.s:2.1}.
\end{proposition}
\begin{IEEEproof}
	See Appendix \ref{a:sp1}
\end{IEEEproof}

The approximation is accurate when the file catalog size $N_f$ is large. As shown in Appendix \ref{a:sp1}, the closed-form solution of $P^*_t$ can be obtained by solving a cubic equation, which is not provided herein for conciseness.

\subsubsection{Offloading Gain}
To evaluate the offloading gain provided by cache-enabled D2D communications, which is characterized by the offloading ratio, both complete transmission and partial transmission should be taken into account.

\begin{proposition}\label{p:2} 
The offloading ratio in the full reuse case is
\begin{equation}
\label{equ.p:2.1}
\begin{aligned}\textstyle
p^a_1(P_t,\rho) =& \textstyle \sum_{i=1}^{N_f} p_r(i) \int_{0}^{r_c}  \frac{f_i(r)}{\ln(1+\Gamma_1)}\int_{0}^{\Gamma_1} \frac{e^{-\phi_i(t,r)} }{1+t}  dt dr,
\end{aligned}
\end{equation}
and $p_1(P_t,\rho) \leq p^a_1(P_t,\rho) \leq p_o$, both equalities will hold if $\rho \rightarrow \infty$ or if $r^{\alpha}\sigma_0^2 \rightarrow 0$ and $\lambda_I \rightarrow 0$.
\end{proposition}
\begin{IEEEproof}
	See Appendix \ref{a:p2} 
\end{IEEEproof}\vspace{2mm}

The first condition $\rho \rightarrow \infty$ means that all helpers have infinite battery capacity, which is the scenario where the user devices  are charging when  acting as the DTs. The second condition $r^{\alpha}\sigma_0^2 \rightarrow 0$ and $\lambda_I \rightarrow 0$ indicate that all interference are eliminated and the SNR is infinite, because $r^{\alpha}\sigma_0^2 = 1/(\frac{P_tr^{-\alpha}}{\sigma^2})$ is the inverse of the receive signal to noise ratio (SNR) at the DR  averaged over fading. In this case, although battery is limited, the data rate can be extremely high to complete all transmission via D2D links. In either condition, the offloading probability, offloading ratio and offloading opportunity are equal.

\subsubsection{Energy costs}
In what follows, we derive the energy cost of a DT for a given transmit power and caching policy, with which we can evaluate the energy cost of a DT with the optimized  transmit power and caching policy.
\begin{proposition}\label{t:2}
The average energy consumed at a DT for a complete transmission is
\begin{equation}
\label{equ.t:2}
\begin{aligned}\textstyle
\bar{E}_1 = \rho V_0 Q - \rho V_0 Q \sum_{i=1}^{N_f} \frac{p_r(i)}{p_1(P_t,\rho)}
\int_{0}^{r_c} f_i(r)\ln(1+\Gamma_1)\int_{0}^{\frac{1}{\ln(1+\Gamma_1)}}  e^{-\phi_i(e^{\frac{1}{t}}-1,r)}  dt dr
\end{aligned}
\end{equation}

\end{proposition}
\begin{IEEEproof}
See Appendix \ref{a:t2}
\end{IEEEproof}\vspace{2mm}

For the satisfied DR, its DT consumes less energy than the allowed battery consumption. For the D2D link with  partial transmission, the DT consumes energy $\rho V_0Q$ to convey part of the file to the DR. Because a percentage $\frac{p_1(P_t,\rho)}{p_o}$ of requested files can be completely conveyed via D2D transmissions, the average energy consumed by a DT can be obtained as
\begin{equation}
\label{equ.E_a1}
\begin{aligned}\textstyle
\bar{E}^a_1 & = \frac{p_1(P_t,\rho)}{p_o}\bar{E}_1 + \left(1-\frac{p_1(P_t,\rho)}{p_o}\right)\rho V_0Q\\
& \textstyle \stackrel{(a)}{=} \rho V_0 Q - \rho V_0 Q \sum_{i=1}^{N_f} \frac{p_r(i)}{p_o}
\int_{0}^{r_c} f_i(r)\ln(1+\Gamma_1)\int_{0}^{\frac{1}{\ln(1+\Gamma_1)}}  e^{-\phi_i(e^{\frac{1}{t}}-1,r)}  dt dr,
\end{aligned}
\end{equation}
where (a) is obtained by substituting \eqref{equ.p_o}, \eqref{equ.p:2.1} and \eqref{equ.t:2}.

To reflect how much energy consumed at a DT by serving as a helper occupies the battery capacity, we
define the \emph{energy cost} as $\bar{e}_1 = \frac{\bar{E}^a_1}{V_0Q}$.

\subsection{Case 2: TDMA}
\label{s:4.2}
By using TDMA, the DT of a randomly scheduled D2D link transmits the requested file to its corresponding DR, while other DTs stay mute. The data rate of each DR is given by
\begin{equation} \label{equ.R_2} \textstyle
R_2(r)  =  \frac{W}{N_a}\log_2(1+\frac{P_thr^{-\alpha}}{\sigma^2}),
\end{equation}
where $N_a = p_o\lambda  S$ is the average number of DRs in a cell and $S$ is the area of the cell. The muting DTs can turn off some circuits to save energy. We call the circuit power consumed by a muting DT as \emph{idle power}, denoted as $P_{c_I}$, which ranges from a few to tens of mW \cite{KV.TWC}. Then, the energy consumed at a DT to transmit a file via the D2D link can be obtained as,
\begin{equation} \label{E_2}
\begin{aligned} \textstyle
E_2(r) & \textstyle =\frac{F}{N_a R_2(r)}\left(\frac{1}{\eta}P_t+P_c \right) + \left( \frac{F}{R_2(r)} - \frac{F}{N_aR_2(r)}\right) P_{c_I}  =\frac{F}{W\log_2(1+\frac{P_thr^{-\alpha}}{\sigma^2})}\left(\frac{1}{\eta}P_t+ P_c^T\right),
\end{aligned}
\end{equation}
where $P_c^T \triangleq P_c + (N_a-1)P_{c_I}$. Note that $E_2(r)$ is not related to the file index $i$ since the received signals are only corrupted by noise, which is different from $E_1(r,i)$.

\subsubsection{Optimal Transmit Power} From the definition of the offloading probability and \eqref{E_2},
it is easy to obtain $p_2(P_t,\rho)$ by letting $\lambda_I\xi_1-\lambda_i^d\xi_2=0$ in \eqref{equ.t:1.1}  as
	\begin{equation}
	\label{equ.t:3}\textstyle
	p_2(P_t,\rho) = \sum_{i=1}^{N_f} p_r(i) \int_{0}^{r_c} f_i(r) e^{-\Gamma_2r^\alpha \sigma_0^2  } dr,
	\end{equation}
	where $\Gamma_2=e^{\frac{F \left(P_t + \eta P_c^T\right)\ln2}{W\rho Q V_0 \eta }} - 1$.
With the growth of both the number of DRs $N_a$ and idle power $P_{c_I}$, the circuit power $P_c^T$ and hence $\Gamma_2$ increase, which results in the reduction of $p_2(P_t,\rho)$.

To maximize the offloading probability for the cache-enabled D2D communications, the transmit power at each DT can be optimized as follows
\begin{equation}
\label{equ.opt3}
\begin{aligned}
\max_{P_t} \,\, & \textstyle p_2(P_t,\rho)\\
s.t.\quad
& \textstyle 0< P_t \leq P_{\max}.
\end{aligned}
\end{equation}

Again, the closed-form expression of $p_2(P_t,\rho)$ is hard to obtain in general. When $\alpha=2$, by using the similar way to derive \eqref{equ.sp:1.1}, we can approximate the offloading probability as
\begin{equation}
\label{equ.s:4.1}
\begin{aligned}\textstyle
p_2(P_t,\rho)  \approx \sum_{i=1}^{N_f} p_r(i) \pi\lambda_i \frac{1 - e^{-(\sigma_0^2\Gamma_2 + \pi\lambda_i)r_c^2}}{\sigma_0^2\Gamma_2+ \pi\lambda_i}.
\end{aligned}
\end{equation}

Despite that the  offloading probability has complicated expression in general cases,  the closed-form solution of the optimal transmit power for all values of $\alpha$ can be found as follows.

\begin{proposition}\label{pt:4}
The optimal transmit power is
	\begin{eqnarray}
	\label{equ.pt:4.1}\textstyle
	P_t^*=
	\begin{cases}\textstyle
	P_{\max}, & \textstyle P_{\max} < \eta P_c^T \left(\sqrt{\frac{1}{a\eta P_c^T }+\frac{1}{4}}-\frac{1}{2}\right) \\
	\eta P_c^T \left(\sqrt{\frac{1}{a\eta P_c^T }+\frac{1}{4}}-\frac{1}{2}\right), & \textstyle\text{otherwise} \\
	\end{cases},
	\end{eqnarray}
	where $a = \frac{F\ln2}{W\rho Q V_0 \eta }$.
\end{proposition}
\begin{IEEEproof}
	See Appendix \ref{a:pt4}
\end{IEEEproof}\vspace{2mm}

%

\subsubsection{Offloading Gain} By considering both the complete and partial transmission, we can obtain the offloading ratio by using a similar method as for the proof of Proposition \ref{p:2} as
\begin{equation}
\label{ap2:1.1}
\begin{aligned}\textstyle
p^a_2(P_t,\rho) =& \textstyle \sum_{i=1}^{N_f} p_r(i) \int_{0}^{r_c}  \frac{f_i(r)}{\ln(1+\Gamma_1)}\int_{0}^{\Gamma_1} \frac{e^{-tr^{\alpha}\sigma_0^2} }{1+t}  dt dr \nonumber\\
=&\textstyle \sum_{i=1}^{N_f} p_r(i) \int_{0}^{r_c}  \frac{f_i(r)}{\ln(1+\Gamma_1)} e^{r^{\alpha}\sigma_0^2} \int_{0}^{\Gamma_1} \frac{e^{-(t+1)r^{\alpha}\sigma_0^2} }{t+1}  d(t+1) dr\\
=& \textstyle \sum_{i=1}^{N_f} p_r(i) \int_{0}^{r_c}  \frac{f_i(r)}{\ln(1+\Gamma_2)} e^{r^\alpha \sigma_0^2  }\left( Ei(-r^\alpha \sigma_0^2 (\Gamma_2+1)) - Ei(-r^\alpha \sigma_0^2) \right)  dr,
\end{aligned}
\end{equation}
where $Ei(x) = \int_{-\infty}^{x} \frac{e^t}{t}dt$ is a frequently-used special function.


\subsubsection{Energy costs} By using the similar derivation as for Proposition \ref{t:2}, we can obtain the average energy consumed at a DT for a complete transmission with given transmit
power and caching policy as
	\begin{equation}
	\label{equ.t:4}
	\begin{aligned}\textstyle
	\bar{E}_2 = \rho V_0Q - \rho V_0 Q \sum_{i=1}^{N_f} \frac{p_r(i)}{p_2(P_t,\rho)}
	\int_{0}^{r_c} f_i(r) \ln(1+\Gamma_2)	\int_{0}^{\frac{1}{\ln(1+\Gamma_2)}} e^{-(e^{\frac{1}{t}}-1)r^\alpha \sigma_0^2 } dt  dr.
	\end{aligned}
	\end{equation}

\vspace{2mm} Since a percentage $\frac{p_2(P_t,\rho)}{p_o}$ of the requested files can be completely conveyed via D2D transmissions, and the DT only consumes energy $\rho V_0Q$ to help the DR for a partial transmission, by considering both complete and partial transmission, the average energy consumption for a DT can be obtained as
\begin{equation}
\label{equ.E_a2}
\begin{aligned}\textstyle
\bar{E}^a_2 & = \frac{p_2(P_t,\rho)}{p_o}\bar{E}_2 + \left(1-\frac{p_2(P_t,\rho)}{p_o}\right)\rho V_0Q\\
&\textstyle \stackrel{(a)}{=}  \rho V_0 Q - \rho V_0 Q \sum_{i=1}^{N_f} \frac{p_r(i)}{p_o}
\int_{0}^{r_c} f_i(r)\ln(1+\Gamma_2)\int_{0}^{\frac{1}{\ln(1+\Gamma_2)}}  e^{-(e^{\frac{1}{t}}-1)r^\alpha \sigma_0^2 }  dt dr,
\end{aligned}
\end{equation}
where (a) is obtained by substituting \eqref{equ.p_o}, \eqref{equ.t:3} and \eqref{equ.t:4}.

Then, the \emph{energy cost} for a DT to transmit a file  is $\bar{e}_2 = \frac{\bar{E}^a_2}{V_0Q}$.

\section{Simulations}
\label{s:5}
In this section, we validate previous analytical results, and  evaluate the offloading gain of the system and the energy cost at each DT via simulations.

We consider a square cell with side length $500$ m. The users' locations follow a PPP with $\lambda = 0.01$, so that in average there is one user in a $10~\text{m}\times 10~\text{m}$ area. The
path-loss model is $37.6+36.8\log_{10}(r)$, where $r$ is the distance of the D2D link \cite{JMY.JSAC}. $W=20$ MHz and
$\sigma^2 = -100 $ dBm, the
maximal transmit power of each DT is $P_{\max}=200 $ mW ($23$ dBm), the circuit power for an active DT is
$P_c=115.9 $ mW, and the power amplifier efficiency is $\eta = 0.5$ \cite{Andreev.JSAC}. The typical idle power for a muting DT with TDMA is $P_{c_I}=25$ mW \cite{KV.TWC}. The operating voltage at a user device is  $V_0 = 4V$ and the battery capacity is $Q = 1800 $ mAh
(typical for current-generation smartphones).
The file catalog contains $N_f=1000$ files, where each file has a size of $30$ Mbytes (a typical video size on the Youtube website \cite{Mo.Mag13}). The parameter of the Zipf distribution $\beta=1$.  This setup is used in the sequel unless otherwise specified.

Besides the optimized caching policy in Proposition 1 (with legend ``Optimal-x''), we also provide the results for \emph{uniform
caching policy} (i.e., each user selects a file from the catalog uniformly, with legend ``Uniform-x'') and \emph{popularity based caching policy} (i.e., each user selects a file from the catalog according to the content popularity, with legend ``Popularity-x'') as the baseline caching policies, where ``x'' is the number of files cached at each user.
%
%

\vspace{-2mm}
\subsection{Optimal Caching Distribution and Offloading Opportunity}

\begin{figure}[!t]
	\centering
	\subfigure[Optimal caching distribution versus file index]{
		\includegraphics[width=0.475\textwidth]{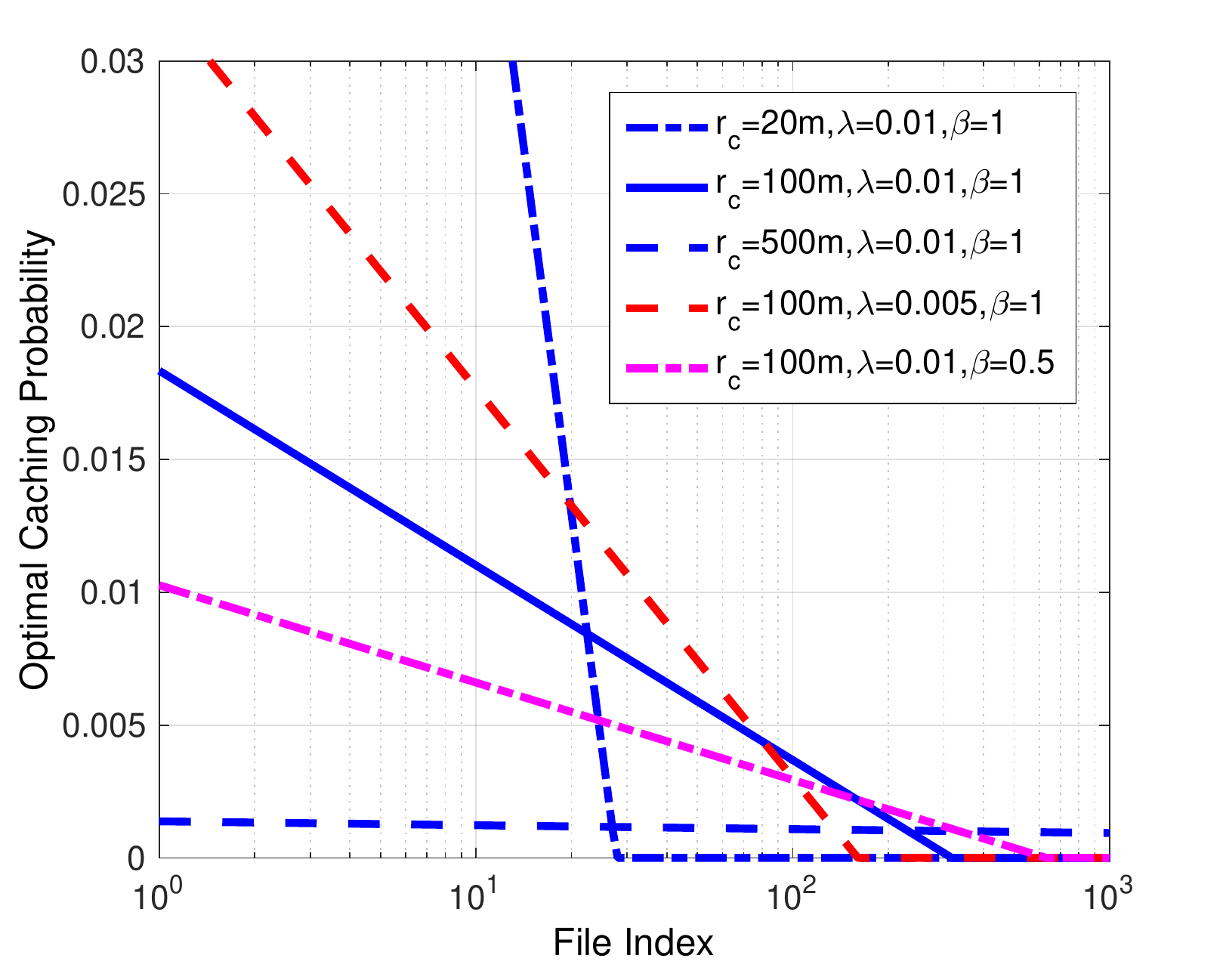}
	}
	\subfigure[Offloading opportunity versus $r_c$]{
		\includegraphics[width=0.475\textwidth]{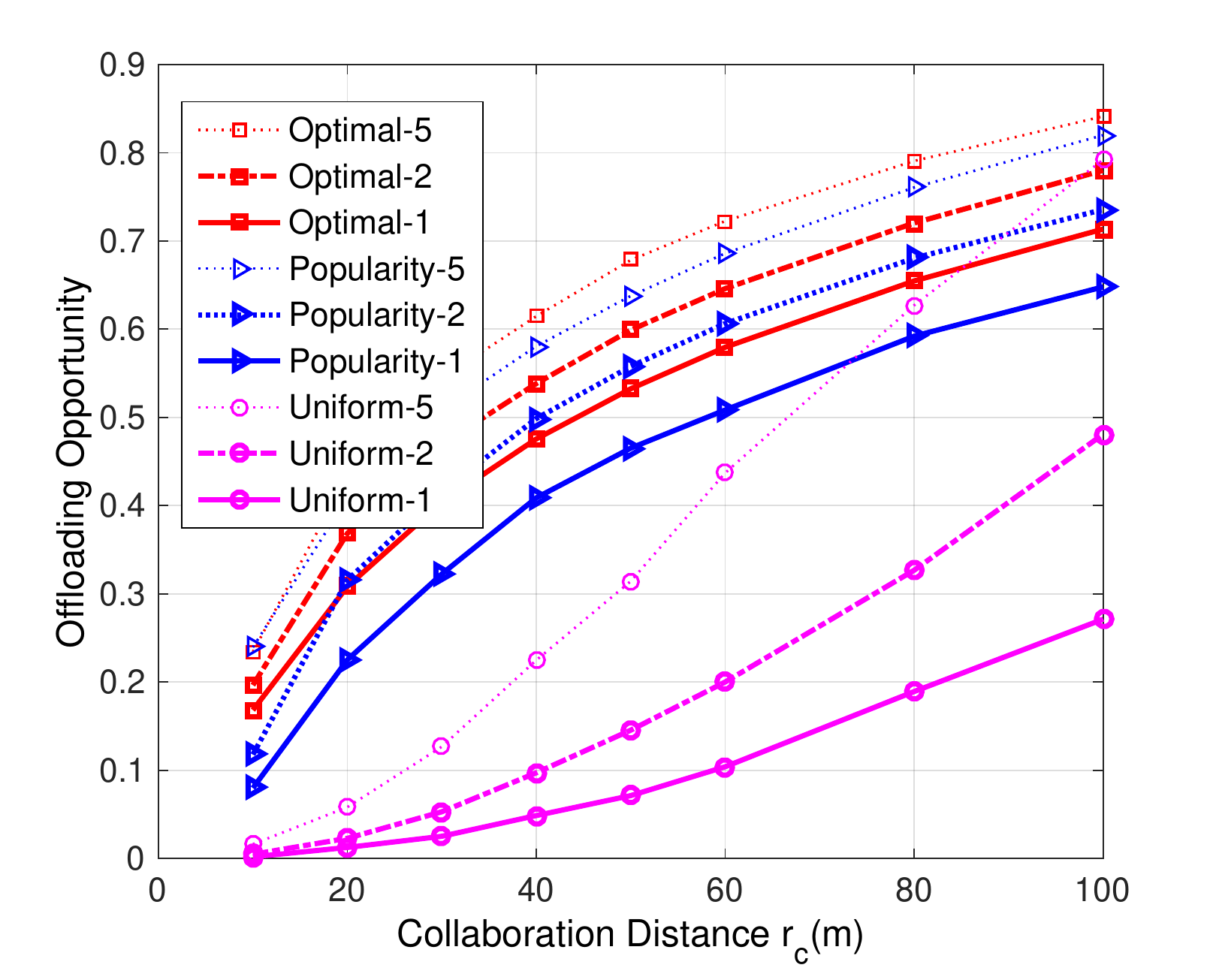}
	}\vspace{-0.20cm}
	\caption{Optimal caching distribution and offloading opportunity with different caching policies.}\label{fig.3}
	\vspace{-0.50cm}
\end{figure}

In Fig. \ref{fig.3}(a), we show the optimal caching distribution for different collaboration distance $r_c$, Zipf parameter $\beta$, and user density $\lambda$. With the increase of $\beta$ and $\lambda$ or decrease of $r_c$, the probability of caching popular files increases, which makes the distribution more ``skewed", and vice versa. When $r_c$ is large enough, say $r_c =500$ m, the caching distribution reduces to a uniform distribution. When $r_c$ is very small, say $r_c=20$ m, the caching distribution makes the probability for caching most popular files very high, which agrees with Proposition 1.

In Fig. \ref{fig.3}(b), we show the simulated offloading opportunity versus the collaboration distance, where each user allows to cache more files. When each user has cached one or two files, the optimized caching policy has the potential to offload more traffic than the popularity based caching policy and even more than the uniform caching policy.
When each user caches more files, the offloading opportunity is improved for all policies, as expected. For large value of $r_c$, say $100$ m, the  offloading opportunities of all caching policies can achieve nearly $0.8$. This indicates that when $r_c$ is large and a user is willing to cache more files, uniform caching policy can also achieve high traffic offloading, despite that it is  not good for D2D throughput in general  \cite{Golrezaei.TWC}.  Nonetheless, the offloading opportunity for caching one file and more files exhibits the same trend, which indicates that caching more files offers essentially the same insight with caching one file, which justifies our assumption in previous analysis.\vspace{-4mm}
\subsection{Validation of Analytical Results}
\begin{figure}[!t]
	\centering
	\subfigure[Offloading performance]{
		\includegraphics[width=0.475\textwidth]{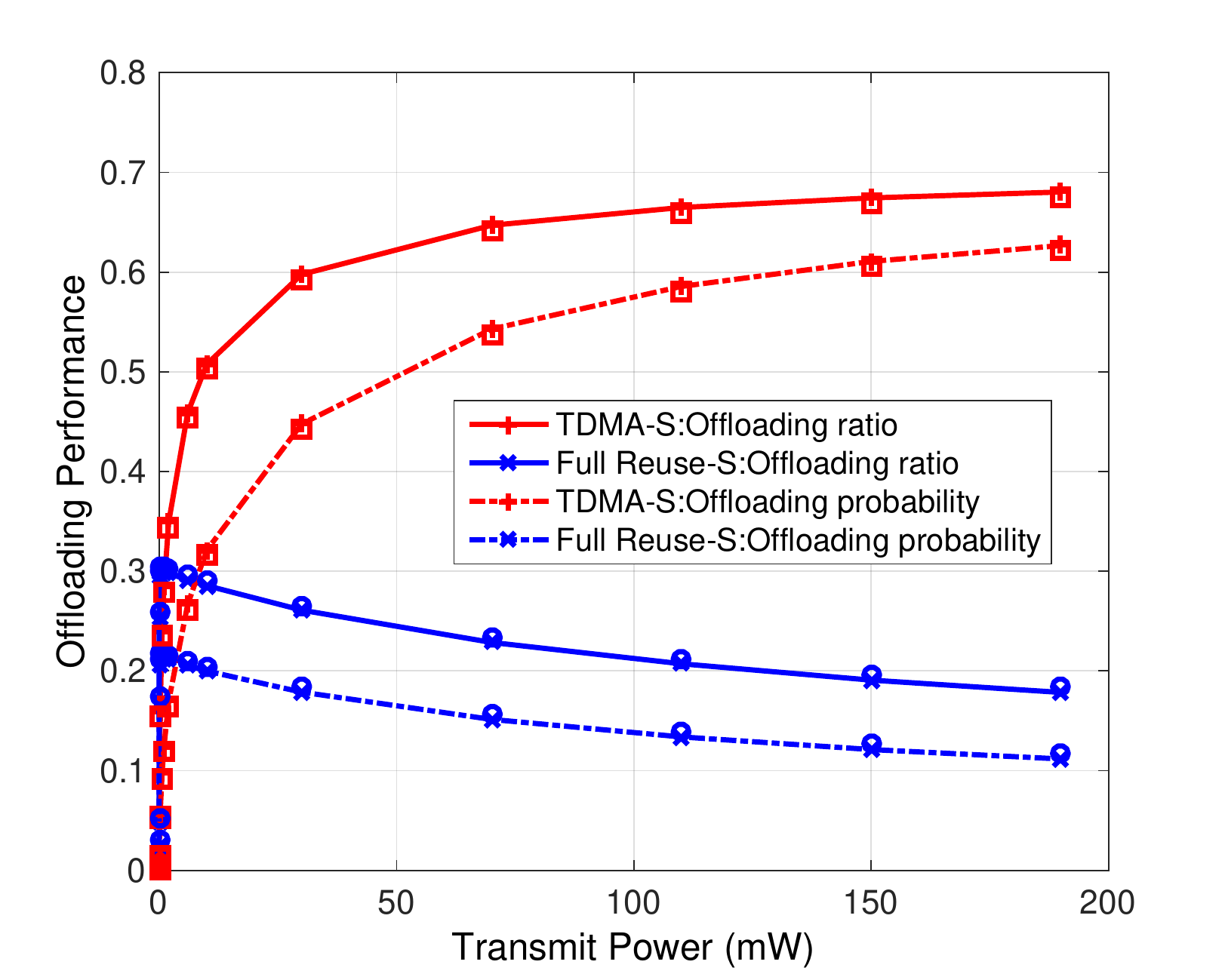}
	}
	\subfigure[Energy cost]{
		\includegraphics[width=0.475\textwidth]{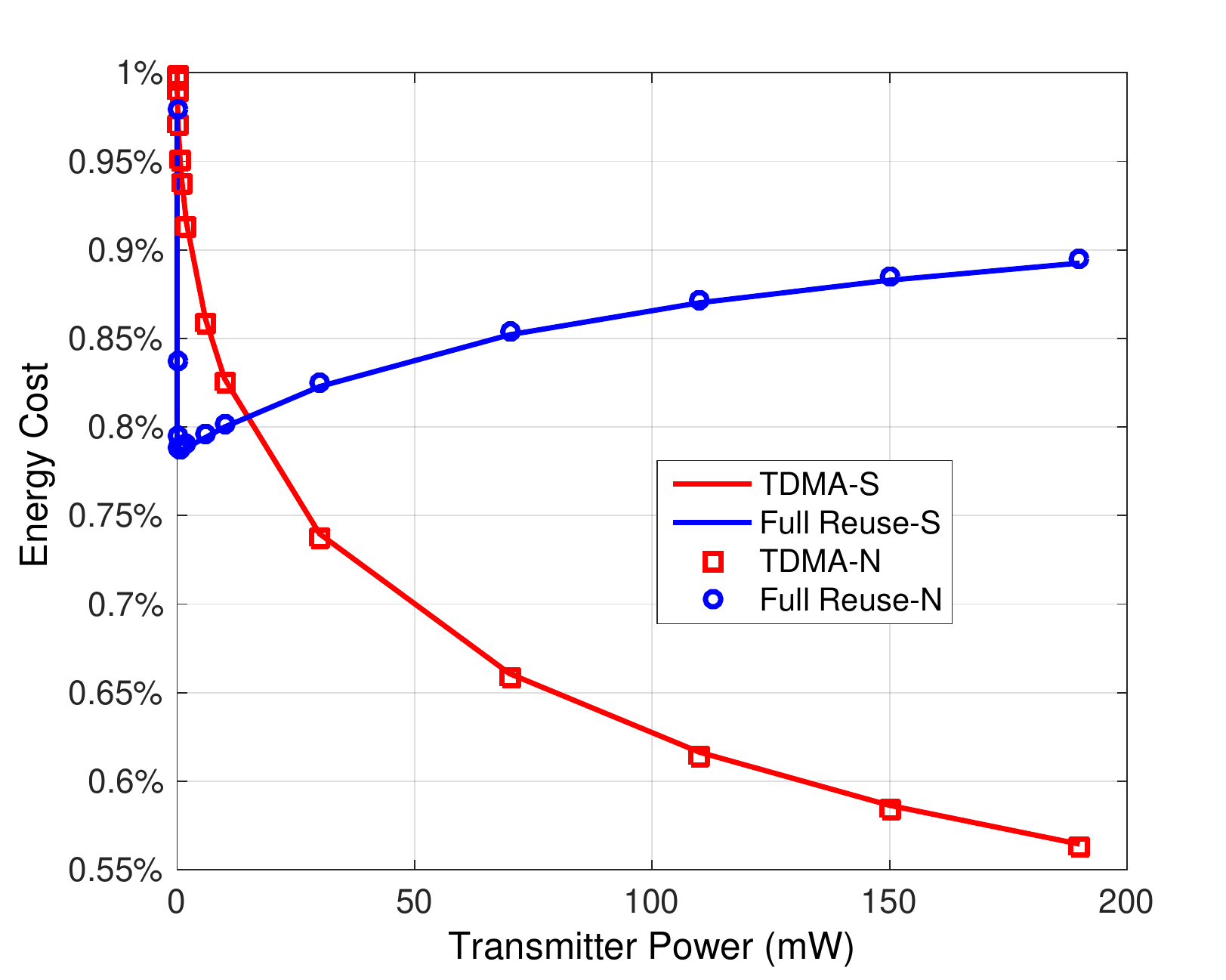}\
	}\vspace{-0.2cm}
	\caption{Validation of the analytical results and show the impact of transmit power $P_t$. $r_c = 100$ m and $\rho=0.01$. S-Simulation results and N-Numerical results. $\bigcirc$ and $\square$ in both Fig. 3(a) and Fig. 3(b) represent the numerical results. }\label{fig.5}
	\vspace{-0.5cm}
\end{figure}

In Fig. \ref{fig.5}, we compare the numerical and simulation results for the offloading probability $p_1(P_t,\rho)$, $p_2(P_t,\rho)$ and offloading ratio  $p^a_1(P_t,\rho)$,  $p_2^a(P_t,\rho)$ respectively for the full reuse and TDMA cases in Fig. \ref{fig.5}(a) and energy cost $\bar{e}_1$ and $\bar{e}_2$ for the two cases in Fig. \ref{fig.5}(b) versus $P_t$. We can see that the numerical results almost overlap with simulation results, which validates our analysis. Moreover,
the trend for $p_1(P_t,\rho)$ and $p_2(P_t,\rho)$ changing with $P_t$ are the same as $p_1^a(P_t,\rho)$ and $p_2^a(P_t,\rho)$, respectively. This suggests that  the optimized transmit power to maximize the offloading probability can also maximize the offloading gain. For the \emph{full reuse case}, $p^a_1(P_t,\rho)$ first increases to achieve the maximal value and then decreases, and $\bar{e}_1$ first decreases and then increases. This is due to the severe interference and the allowed battery consumption. Comparing Fig. \ref{fig.5}(a) and Fig. \ref{fig.5}(b), we can observe that the optimal value of $P_t$ to maximize the offloading probability $p_1^a(P_t,\rho)$ can nearly minimize the energy cost $\bar{e}_1$. This is because to maximize the offloading ratio the transmit power should be reduced in an interference environment and then more DTs consume less power than the allowed battery consumption. For the \emph{TDMA case}  in the considered setting, increasing $P_t$ can always improve $p_2^a(P_t,\rho)$. Moreover, $\bar{e}_2$ always decreases, because increasing $P_t$ can  shorten the duration of transmission and hence can reduce the circuit power consumption.

\begin{figure}[!b]
	\centering
	\subfigure[Offloading probability, $\alpha=2$]{
		\includegraphics[width=0.475\textwidth]{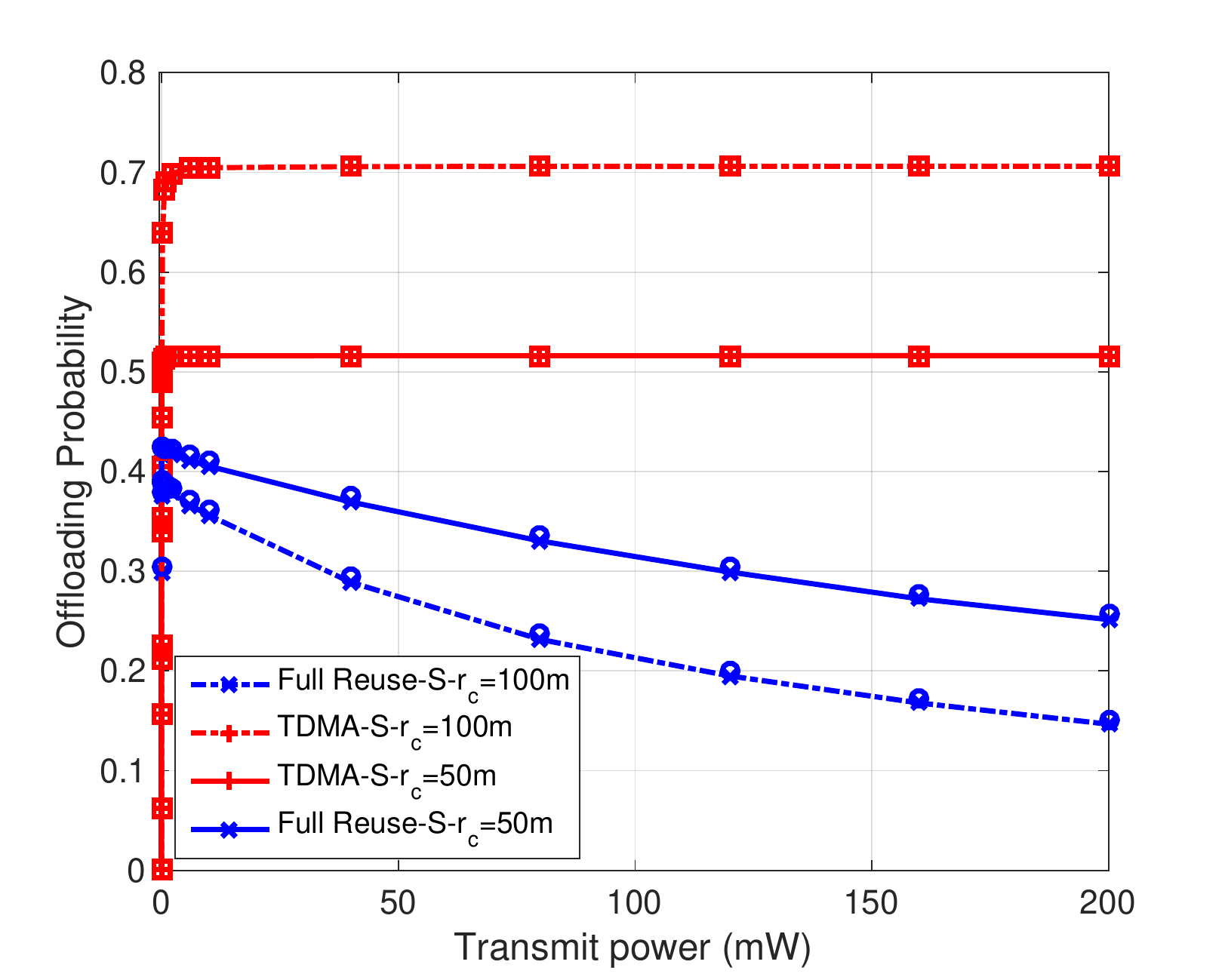}
	}
	\subfigure[Offloading probability, $\alpha=4$]{
		\includegraphics[width=0.475\textwidth]{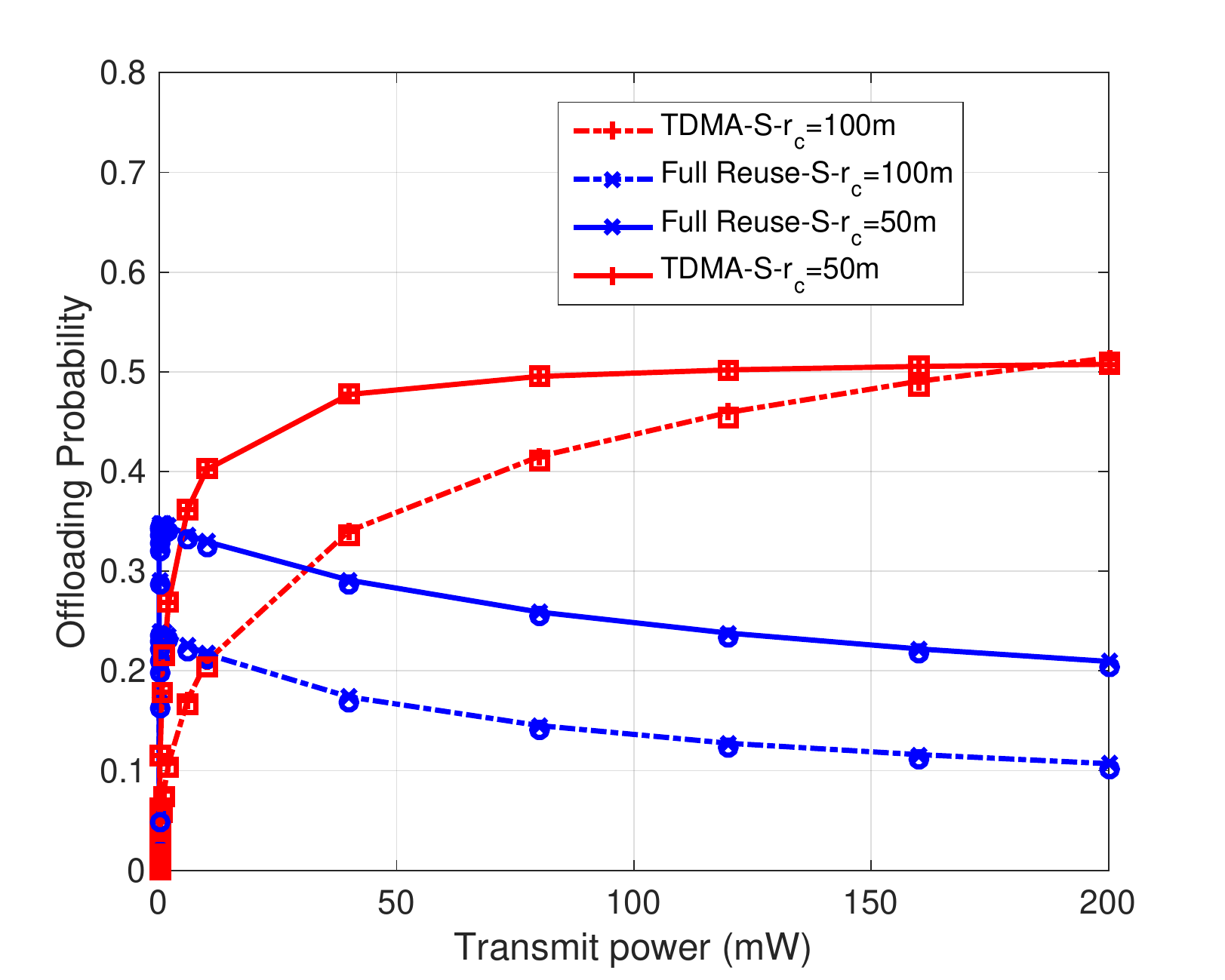}\
	}\vspace{-0.2cm}
	\caption{Validation of the analytical results in special cases. $\rho=0.01$. S-Simulation results. $\bigcirc$ and $\square$ represent the numerical results.  For $\alpha=2$, the interference generated by the DTs far away from $100$ m is ignored in analytical results. }\label{fig.5a}
	\vspace{-0.5cm}
\end{figure}
In Fig. \ref{fig.5a}, we compare the numerical and simulation results for the offloading probability under special channel models. By using the same approach as for deriving the closed form solution for the LOS channel, we can also derive the closed-form expressions for the offloading probability under $\alpha=4$, which are not shown for conciseness. We can see that the numerical results almost overlap with the simulation results, which indicates that the approximations in \eqref{equ.sp:1.1} and \eqref{equ.s:4.1} are accurate. When $\alpha=2$ and $\alpha=4$, the offloading probability for the full reuse case first increases and then decreases with $P_t$, and the offloading probability for TDMA case always grows with $P_t$ in the considered setting, which are the same as Fig. \ref{fig.5}.  It implies that the optimal solution of $P_t$ can be found by bisection searching efficiently in general channels, because  $2 \le \alpha \le 4$ in practical channels among D2D links \cite{JMY.JSAC}.

\begin{figure}[!b]
	\centering
	\subfigure[Offloading probability]{
		\includegraphics[width=0.475\textwidth]{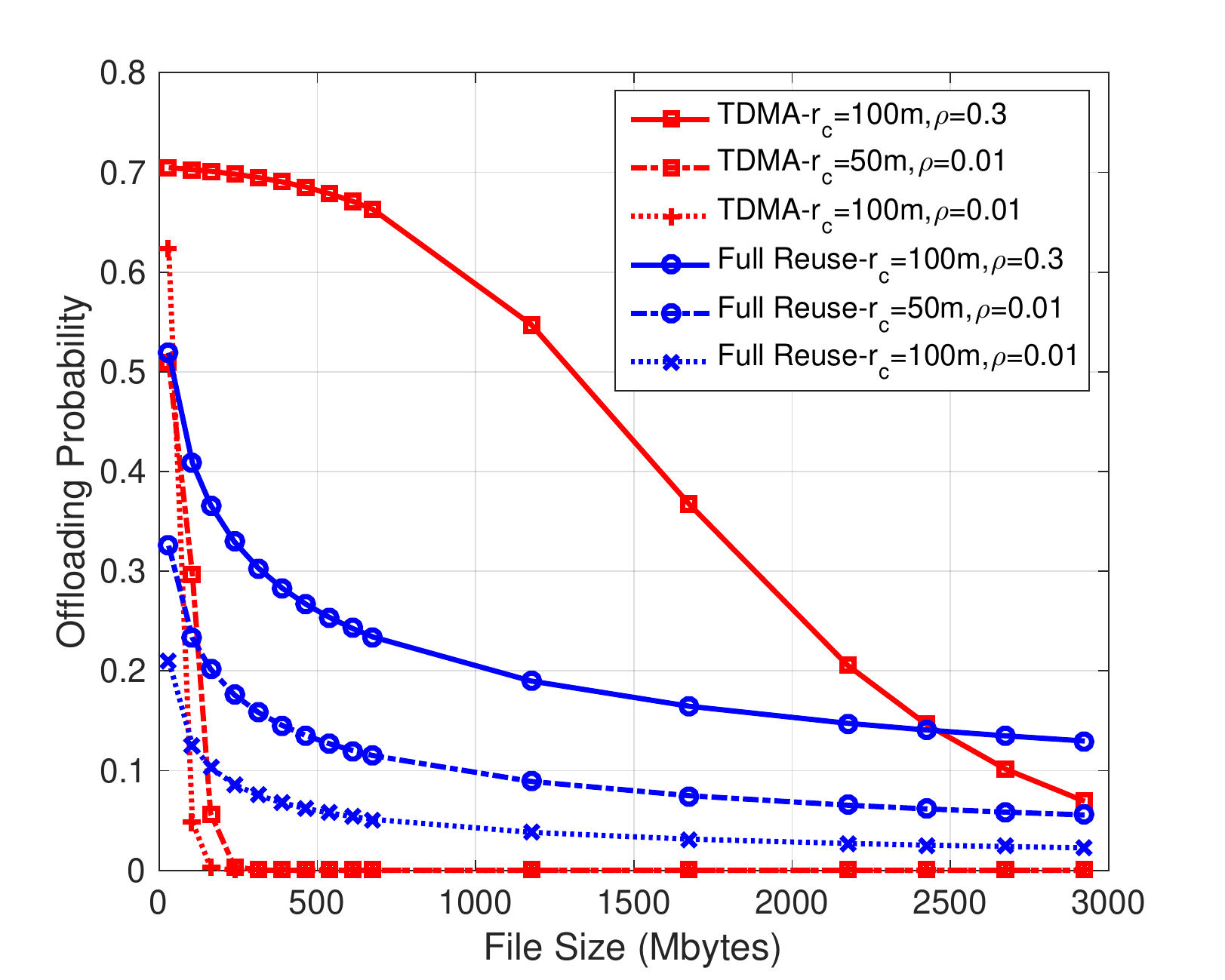}
	}
	\subfigure[Optimal transmit power]{
		\includegraphics[width=0.475\textwidth]{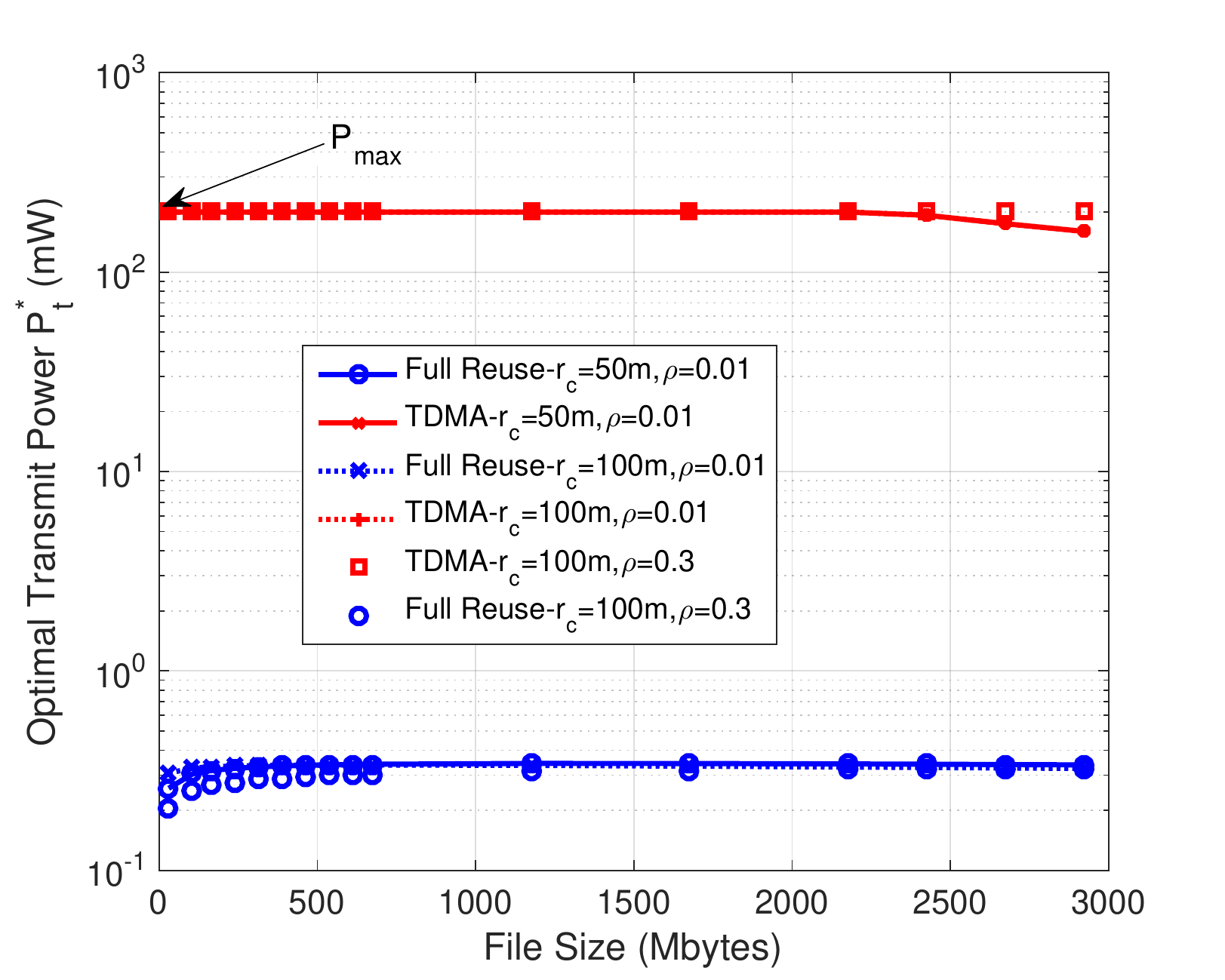}\
	}\vspace{-0.2cm}
	\caption{Impact of file size $F$ on the offloading probability and optimal transmit power. }\label{fig.f}
	\vspace{-0.5cm}
\end{figure}

\subsection{Impacts of Key Parameters and Self Offloading}
In what follows, we analyze the impact of the file size, idle power, content popularity, energy consumption allowed by user device, as well as the self offloading on the offloading gain and energy cost for full reuse and TDMA cases with numerical results. The optimized transmit power and optimized caching policy are used, unless otherwise specified.

In Fig. \ref{fig.f}, we show the impact of file size. We can see that for TDMA, $P^*_t \neq P_{\max}$ only when $\rho=0.01$ and $F>2.2$ GBytes. In all other cases, transmitting with $P_{\max}$ is optimal for TDMA. We can also observe that with the growth of $F$, the offloading probability for both full reuse and TDMA cases decreases. This implies that cache-enabled D2D communications is more applicable for offloading traffic of delivering small size files.

\begin{figure}[!t]
	\centering
	\subfigure[Offloading ratio]{
		\includegraphics[width=0.475\textwidth]{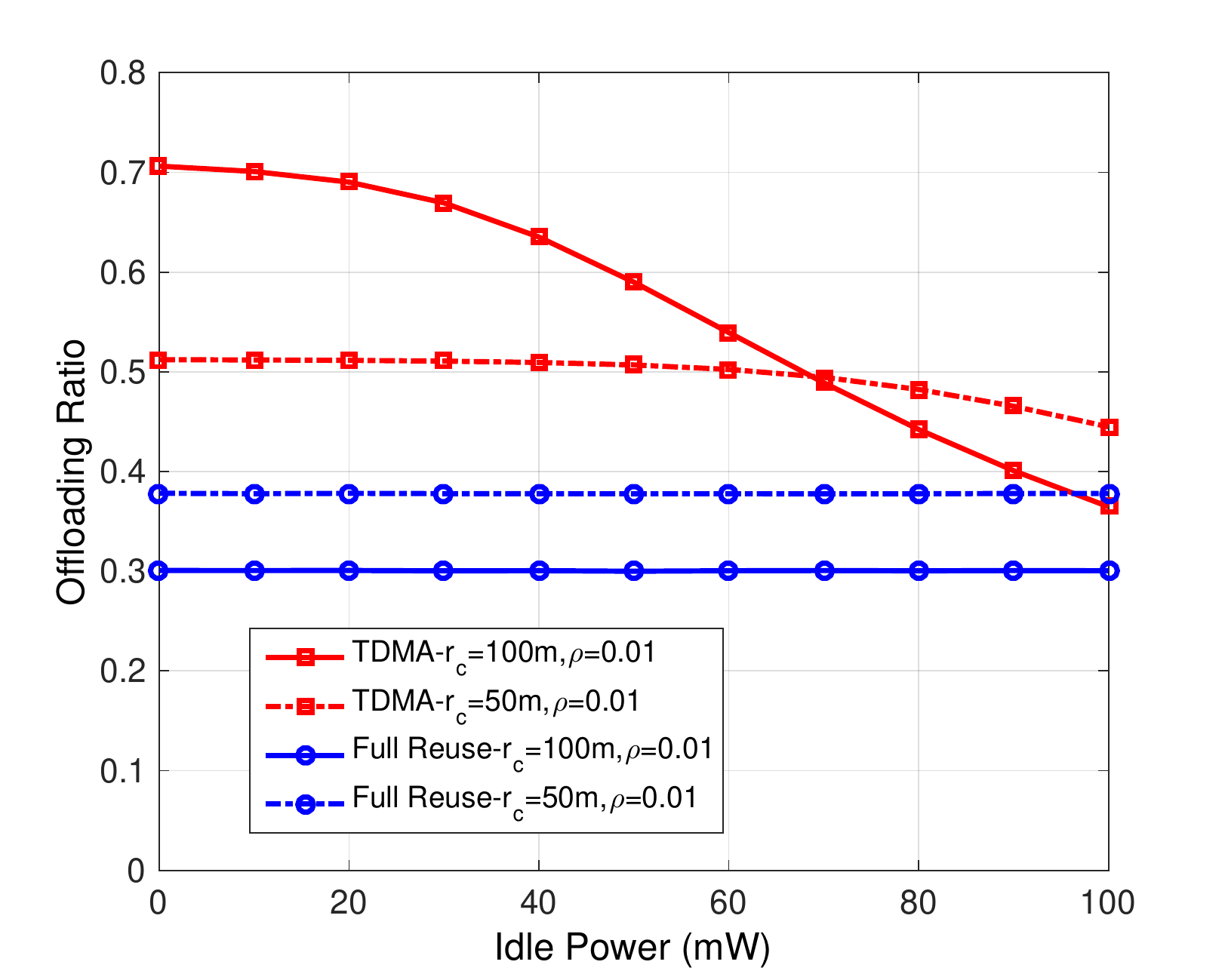}
	}
	\subfigure[Energy cost]{
		\includegraphics[width=0.475\textwidth]{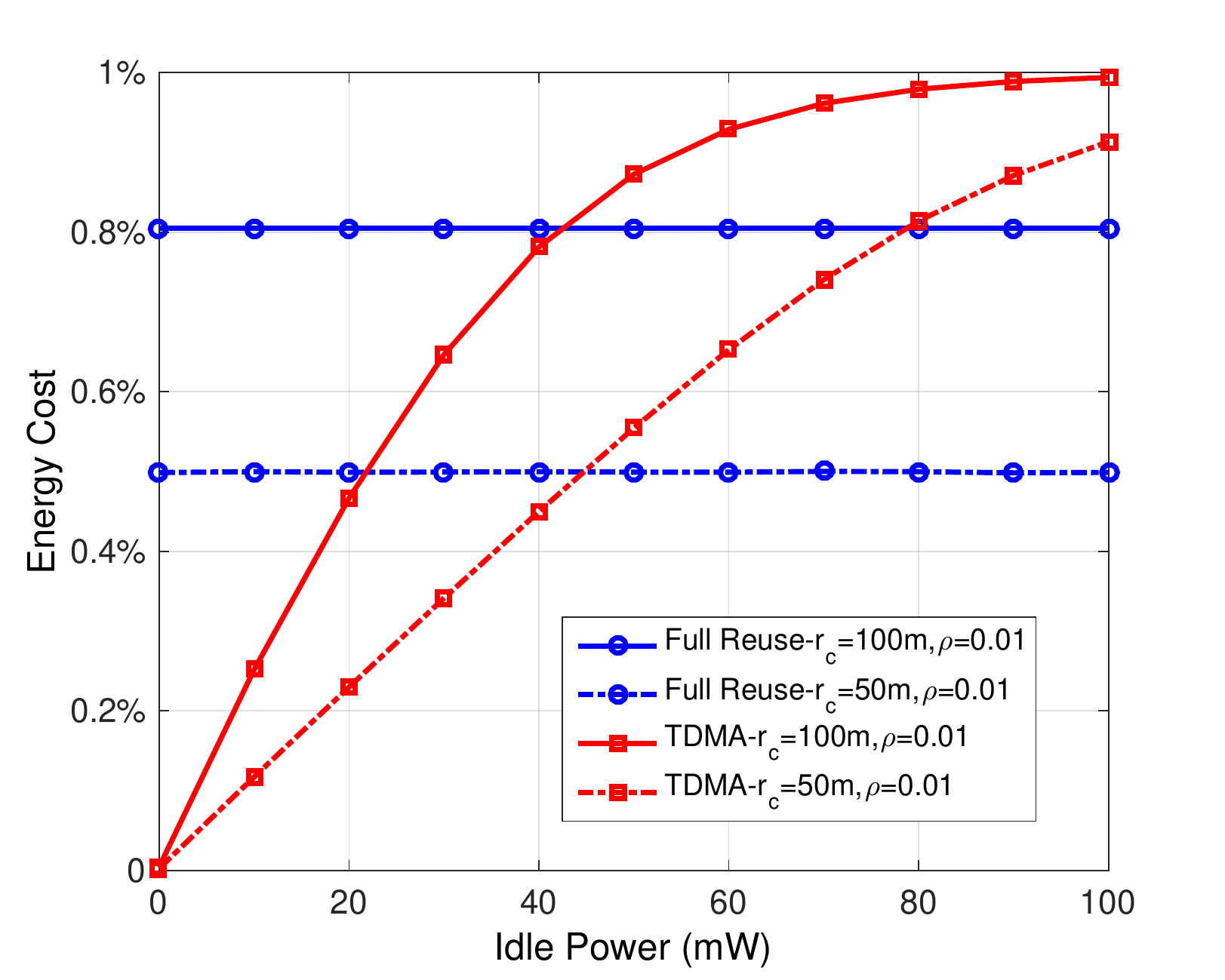}
	}\vspace{-0.2cm}
	\caption{Impact of the idle power of a muting DT under TDMA}\label{fig.C}
	\vspace{-0.50cm}
\end{figure}

  \begin{figure}[!b]
  	\centering
  	\subfigure[Offloading ratio]{
  		\includegraphics[width=0.475\textwidth]{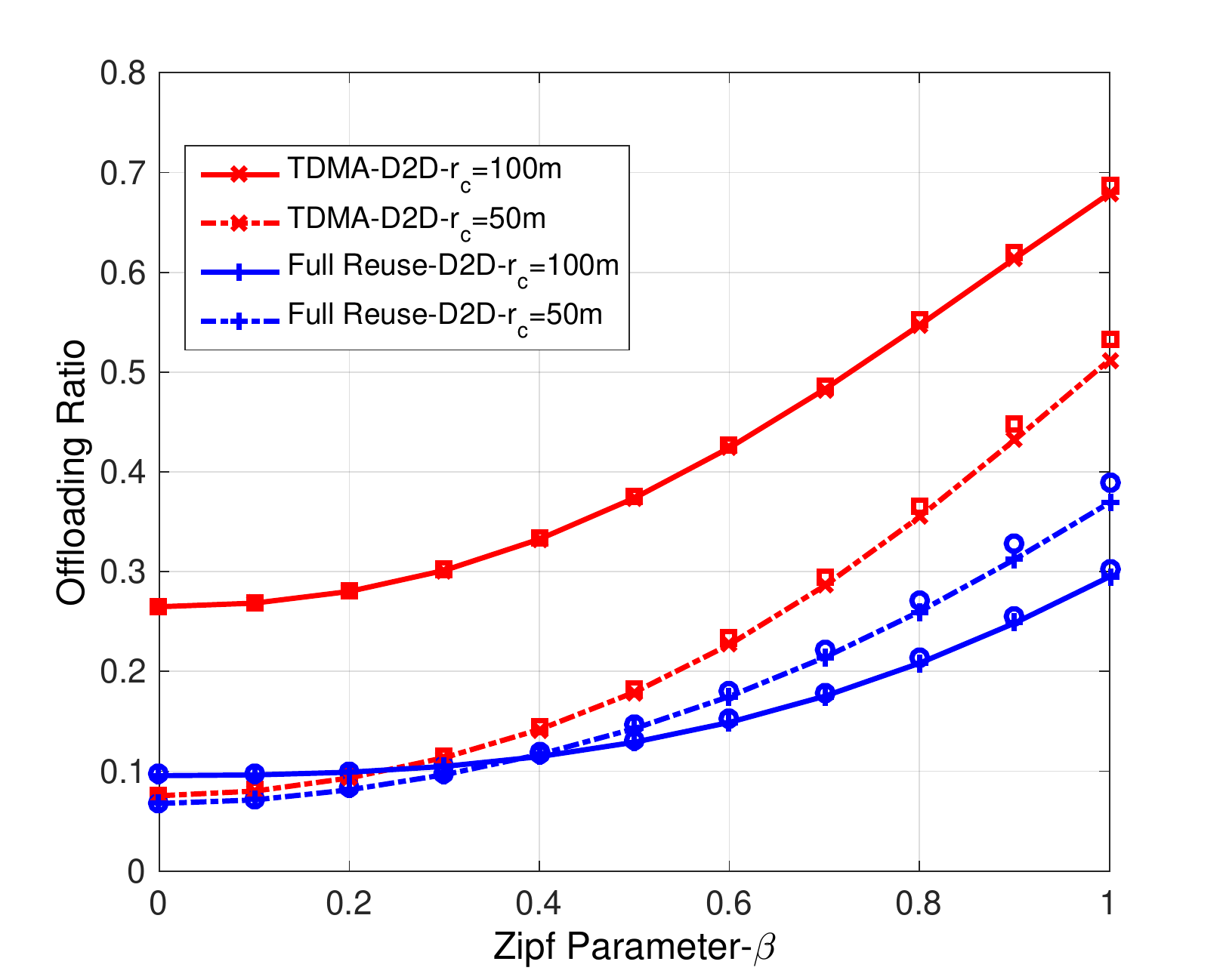}
  	}
  	\subfigure[Energy cost]{
  		\includegraphics[width=0.475\textwidth  ]{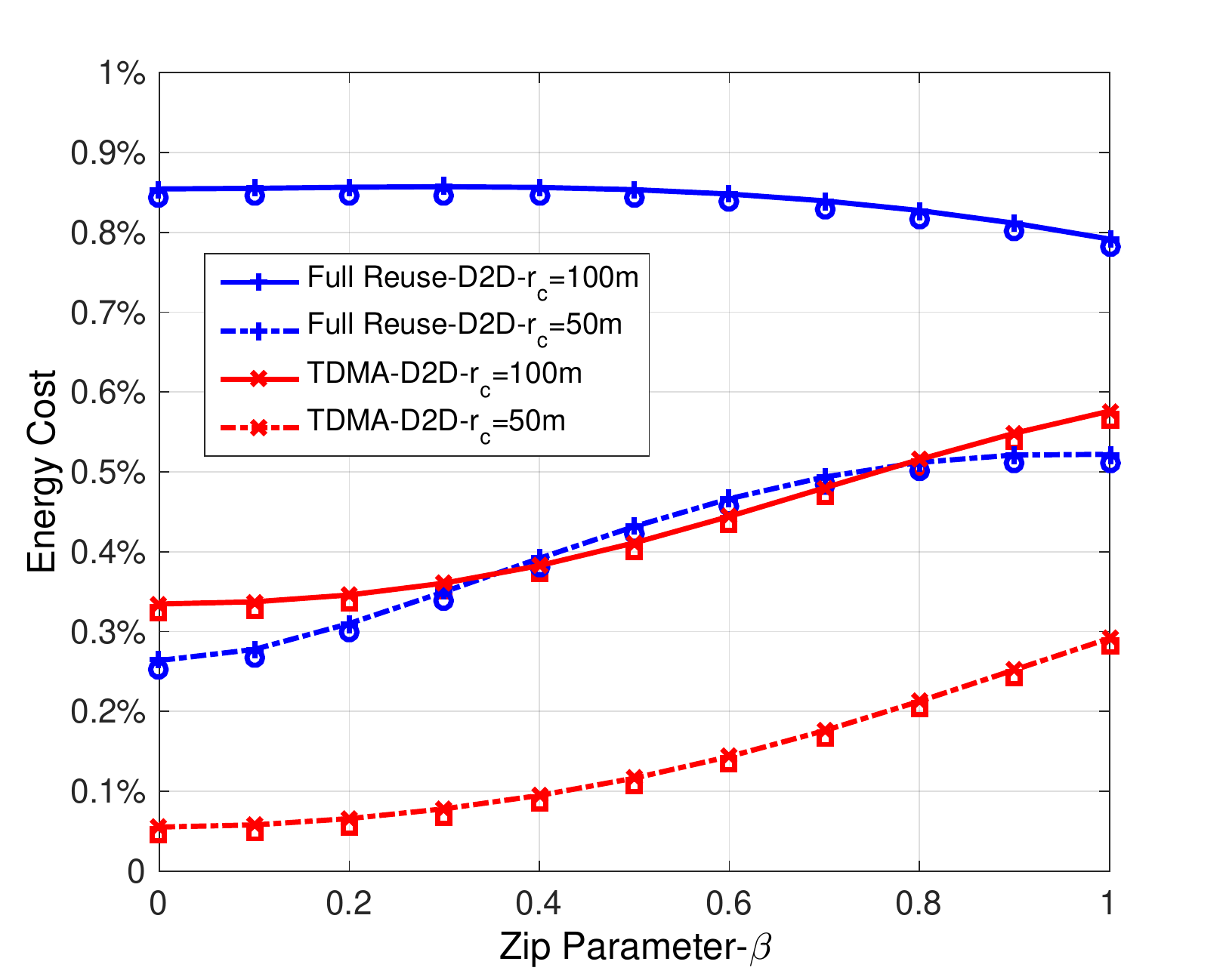}
  	}\vspace{-0.2cm}
  	\caption{Impact of content popularity and self offloading. $\bigcirc$ and $\square$ represent taking self offloading into account, $\rho=0.01$.}\label{fig.7}
  	\vspace{-0.60cm}
  \end{figure}

In Fig. \ref{fig.C}, we show the impact of the idle power $P_{c_I}$. We can see that the offloading ratio of TDMA is always larger than that of the full reuse.  The increase of $P_{c_I}$ directly leads to higher energy cost at each DT with TDMA. When only 1\% battery capacity can be used and $P_{c_I} > 40$ mW, the energy cost at a DT with TDMA is larger than full reuse. When $\rho =30$\% (not shown in the figure), the energy cost of full reuse is around 12\%, which is much larger than the energy cost  of TDMA that changes from 0 to 2.5 \% with the increase of $P_{c_I}$.
\begin{figure}[!b]
	\centering
	\subfigure[Offloading ratio]{
		\includegraphics[width=0.475\textwidth  ]{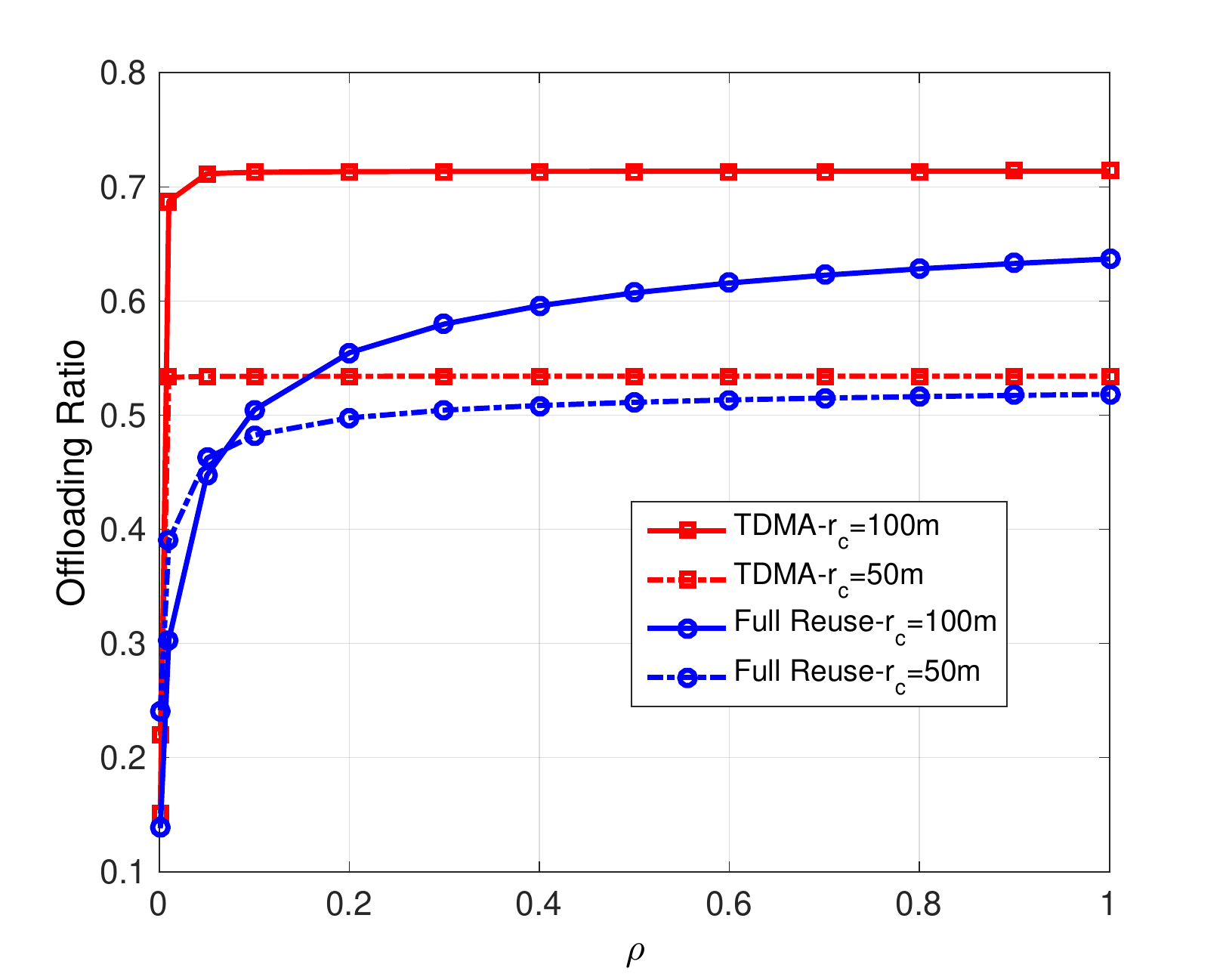}
	}
	\subfigure[Energy cost]{
		\includegraphics[width=0.475\textwidth  ]{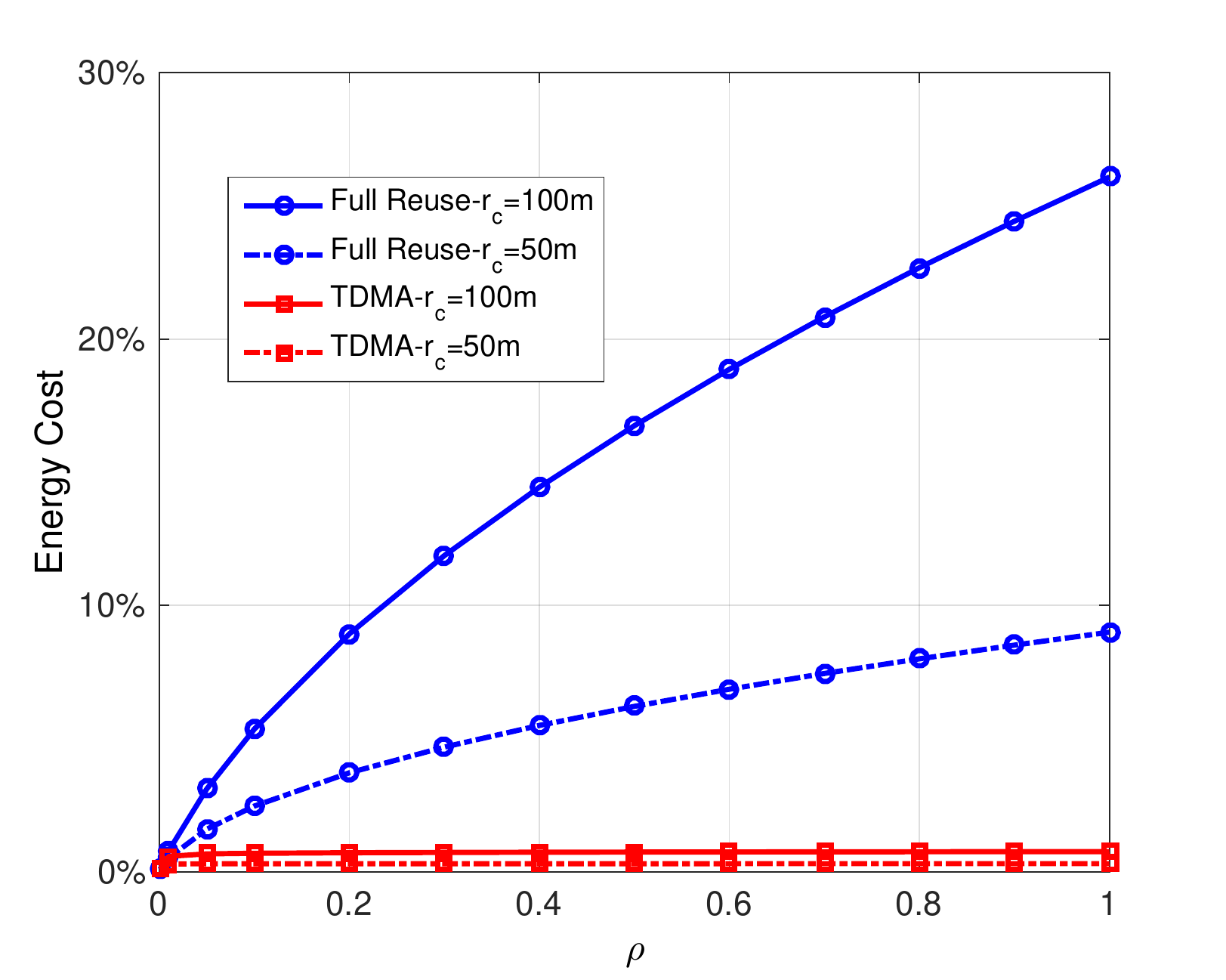}
	}\vspace{-0.2cm}
	\caption{Impact of the allowed fraction of battery consumption $\rho$.}\label{fig.7a}
	\vspace{-0.60cm}
\end{figure}

In Fig. \ref{fig.7}, we show the impact of the Zipf parameter $\beta$ and self offloading. As expected, the offloading ratio increases rapidly with $\beta$ due to the high cache hit rate. The energy cost  increases with $\beta$ for TDMA, but first increase and finally decreases with $\beta$ for full reuse. This can be explained as follows. On one hand, larger $\beta$ leads to smaller D2D link distance, which reduces the energy cost. On the other hand, larger $\beta$ leads to more DTs, which generates  more severe interference for full reuse and longer muting time for TDMA, both of which increase the energy cost. Since the reduction in D2D link distance is dominant for full reuse, the energy cost finally decreases. Besides, the offloading ratio including both cache-assisted D2D communications and self offloading is larger than that only contributed by  the cache-enabled D2D, and the energy cost including  both is less than that only considering D2D. However, the contribution of self offloading on the performance is marginal.

In Fig. \ref{fig.7a}, we show the impact of the allowed battery consumption $\rho$. We can see that the offloading ratio first increases rapidly and then slowly with $\rho$, whereas the energy cost increases with $\rho$ but is always much less than the allowed battery consumption. This is because for D2D links with better channel state, the DTs can transmit complete files to corresponding DRs with less than $\rho QV_0$ of energy. The results suggest that choosing a proper $\rho$ is important  for operators to balance  benefits (e.g., the offloading gain) and costs (e.g., rewarding users for a larger value of $\rho$). We can also observe that the energy cost for the full reuse case grows more rapidly than the TDMA case, because there are more partial transmission links in the full reuse case than the TDMA case, whose energy consumed by each DT equals the allowed battery consumption.

\begin{figure}[!t]
	\centering
	\subfigure[Changing $r_c$ from 10 m to 400 m]{
		\includegraphics[width=0.475\textwidth]{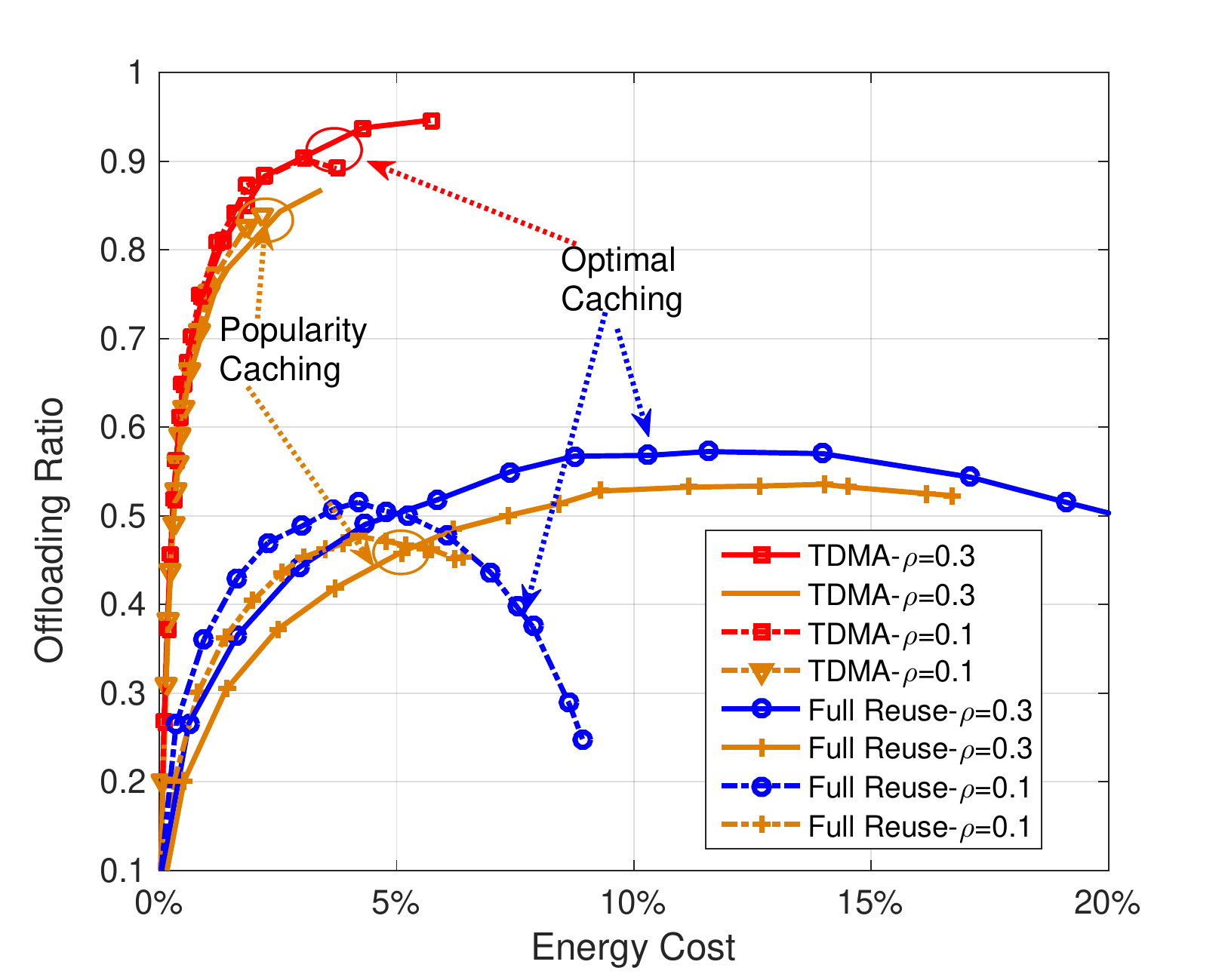}
	}
	\subfigure[Changing $\rho$ from 0 to 1, $r_c=r_c^*$.]{
		\includegraphics[width=0.475\textwidth]{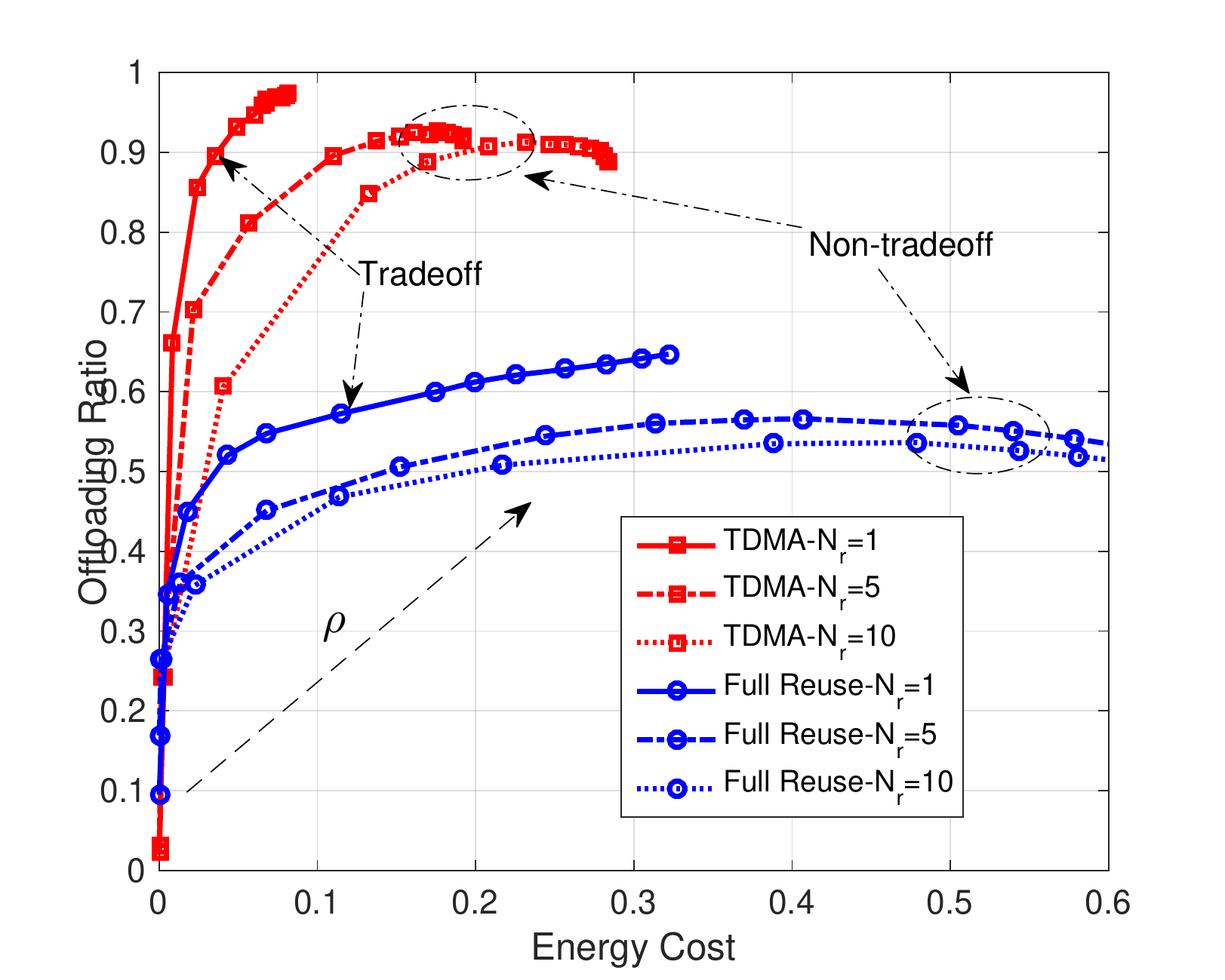}
	}\vspace{-0.2cm}
	\caption{ Relationship between offloading gain and energy cost, $\lambda=0.01$.  $N_r$ is the number of requests each user sends.}\label{fig.8}\vspace{-0.6cm}
\end{figure}
\vspace{-0.2cm}\subsection{Relationships Between Offloading Gain and Energy Cost}
In the sequel, we show the relation between offloading ratio and energy cost with the optimal transmit power and optimal caching policy.

In Fig. \ref{fig.8}(a), the offloading ratio is adjusted by changing the collaboration distance $r_c$ from $10$ m to $400$ m, where popularity based policy is also simulated for comparison. We can observe an optimal $r_c$ to maximize the offloading ratio for a given $\rho$, which are 350 m, 300 m, 100 m, 80 m, respectively for ``TDMA, $\rho$ = 0.3", ``TDMA, $\rho$ = 0.1", ``Full Reuse, $\rho$ = 0.3", and  ``Full Reuse, $\rho$ = 0.1" with the optimal caching policy. This is because the full reuse scheme is interference limited and the TDMA scheme is transmit power limited. With the growth of $r_c$, the average D2D communication distance increases, and hence the energy cost increases, whereas the very limited battery consumption  allowed for helping others makes the offloading ratio decrease. Compared with popularity based caching policy, the optimal caching policy can improve the offloading ratio and reduce energy cost.

In Fig. \ref{fig.8}(b), the offloading ratio is adjusted by changing $\rho$ from $0$ to $1$, where the optimal $r_c$ maximizing the offloading gain and optimal caching policy is used. To show what happens if a helper serves multiple requests, here each user sends $N_r$ requests sequentially according to the Zipf distribution. As a result, the helper that cached the most popular files may be requested multiple times and serve as a DT for multiple users. When  $N_r$ = 1, there exists a tradeoff between offloading gain and energy cost. When $N_r > 1$, a large energy cost may not yield a high offloading gain. This is because with larger $\rho$, a DT will consume more energy before interrupting the transmission for a D2D link with bad channel condition, and will soon run out of battery for serving subsequent requests. Consequently, each DT can serve fewer requests, which leads to the reduction of the offloading gain. Nonetheless, it is interesting to observe that the energy cost to support high offloading ratio  is low. Even when $N_r$ = 10, to offload around $80\%$ of traffic, the average energy consumption at each DT with TDMA only consumes around $10\%$ battery capacity. This suggests that cache-enabled D2D communications is cost-efficient for offloading by optimizing the collaboration distance and selecting a proper transmission scheme.

\vspace{2mm}In the following, we provide a brief summary of the simulation results.
\begin{itemize}
	\item \emph{Caching policy}: When the collaboration distance is small or only one file is cached, optimizing caching policy can improve offloading gain and reduce energy cost.
	\item \emph{Transmission scheme}: When the file size is not large, TDMA is superior to full reuse with typical value of idle power. The optimization of transmit power to maximize the offloading probability also helps increase the offloading gain and reduce the energy cost. For the TDMA case, the DT can simply transmit with $P_{\max}$ to maximize the offloading gain if the file size is not too large. For the full reuse case, optimizing transmit power is important.
	\item \emph{Parameter setting}: There exists an optimal value of $r_c$ that maximizes the offloading gain for a given $\rho$. When each DT only serves one request, both offloading gain and energy cost increase with $\rho$. When each DT could serve multiple requests, a large value of $\rho$ not only causes large energy cost but also reduces the offloading gain.
	\item \emph{Gain and costs}: When the file size is not very large, a high offloading gain can be achieved by a low energy cost if the collaboration distance, transmission scheme and caching policy are judiciously designed and the  value of $\rho$ is properly selected.
\end{itemize}


\section{Conclusion}
\label{s:6}
In this paper, we quantified the offloading gain of cache-enabled D2D communications after taking the user allowed battery consumption into account and evaluated the energy consumed at a helper user. We considered a user-centric caching and transmission protocol, where the energy consumed for transmission can be controlled by a collaboration distance. We first optimized a proactive caching policy with given collaboration distance, with which the offloading opportunity can be maximized. For either full reuse or TDMA (round-robin) scheduling, we then optimized the transmit power to deliver a file via D2D link, where the percentage of satisfied users is maximized. With the optimized probabilistic caching policy and optimized transmit power, we evaluated the offloading gain of the system and the energy cost of a D2D transmitter, and investigated their relationship. Simulation results showed that high offloading gain can be obtained in practice by cache-enabled D2D with low energy cost at each help user, if the collaboration distance, transmission scheme and caching policy are optimized and the allowed battery consumed by each D2D transmitter for conveying one file is properly set.

\appendices
\numberwithin{equation}{section}
\vspace{-0.2cm}
\section{Proof of Proposition \ref{p:1}}
\label{a:p1} 
Denote
$x_i \triangleq \frac{\ln(p_r(i))}{\lambda \pi r_{c}^2}$ and $ v \triangleq \frac{1}{\lambda \pi r_{c}^2}\ln(\frac{-\mu}{\pi \lambda r_{c}^2})$.
Then, considering  $\sum^{N_f}_{i=1} p_c(i) =1$ and from (\ref{equ.Lag_3}) we have
\begin{equation} \label{equ.proof_1.1} \textstyle
\sum_{i=1}^{N_f} [x_i-v]^+ =1.
\end{equation}

Since problem \eqref{equ.opt1} is convex, the solution of $v$  found from this necessary  condition is globally optimal, and with it the optimal caching distribution can be obtained.

As shown in \eqref{equ.Lag_3},  $p^*_c(i)$ decreases when $p_r(i)$ decreases. As shown in (\ref{equ.p_r}),
$p_r(i)$ is a decreasing function of file index $i$. This indicates that $p^*_c(i)$ is a decreasing function of $i$. Thus, there exists a unique  file index $i^*\leq N_f$, with which
$p^*_c(i)>0$ if $i \le i^*$, $p^*_c(i)= 0$ otherwise. As a result, finding the solution of $v$ from  \eqref{equ.proof_1.1} is equivalent to finding the index $i^*$ from $\sum_{i=1}^{i^*} (x_i-v) =1$. Once $i^*$ is found, the solution of \eqref{equ.proof_1.1} can be obtained as
\begin{equation} \label{equ.proof_1.3}\textstyle
v^* = \frac{\sum_{i=1}^{i^*}x_i - 1}{i^*}.
\end{equation}

\emph{Case 1}: When $i^*=N_f$, from \eqref{equ.Lag_3} and $p_c(i)>0$ we have $p^*_c(N_f)=x_{N_f}-v > 0$, which can be rewritten as $\sum_{i=1}^{N_f}(x_i - x_{N_f})
    < 1$ after substituting $v$ in (\ref{equ.proof_1.3}), then
\begin{equation} \label{equ.proof_1.4}
\begin{split}\textstyle
\sum_{i=1}^{N_f}(x_i - x_{N_f}) = \sum_{i=1}^{N_f} \frac{\ln(p_r(i))-\ln(p_r(N_f))}{\lambda \pi r_{c}^2} {=} \frac{\beta}{\lambda \pi r_{c}^2} \sum_{i=1}^{N_f} \ln(N_f/i) = \frac{\beta}{\lambda \pi r_{c}^2}\ln(\frac{N_f^{N_f}}{N_f!})< 1,
\end{split}
\end{equation}
which can be rewritten as $\frac{(N_f)^{N_f}}{N_f!} < e^{\frac{\lambda \pi r_{c}^2}{\beta}}$. By substituting $v^*$ in (\ref{equ.proof_1.3}) into (\ref{equ.Lag_3}), the optimal caching distribution can be
derived as
\begin{equation} \label{equ.proof_1.5}\textstyle
p^*_c(i) = \frac{\beta}{\lambda \pi r_{c}^2 N_f}\sum_{j=1}^{N_f}\ln(\frac{j}{i})+\frac{1}{N_f}.
\end{equation}

\emph{Case 2}:  When $i^*<N_f$, $p^*_c(i^*) = x_{i^*} - v >0$ and $x_{i^*+1} - v \leq 0$. By
    substituting $v$ in (\ref{equ.proof_1.3}) into these two inequalities, we have $
\sum_{i=1}^{i^*}(x_i - x_{i^*+1}) \geq 1$ and $\sum_{i=1}^{i^*}(x_i - x_{i^*}) < 1$,
which can be further derived by substituting $p_r(i)$ in (\ref{equ.p_r}) and $x_i$ as
\begin{equation} \label{equ.proof_1.7}
\begin{split}\textstyle
\frac{\beta}{\lambda \pi r_{c}^2}\ln(\frac{(i^*+1)^{i^*}}{i^*!})\geq 1, \quad \frac{\beta}{\lambda \pi r_{c}^2}\ln(\frac{(i^*)^{i^*}}{i^*!}) < 1.
\end{split}
\end{equation}
Then, $i^*$ satisfies $\frac{(i^*+1)^{i^*}}{i^*!} \geq e^{\frac{\lambda \pi r_{c}^2}{\beta}}$ and
$\frac{(i^*)^{i^*}}{i^*!} < e^{\frac{\lambda \pi r_{c}^2}{\beta}}$.
With Stirling formula \cite{SF.Approx}, $\sqrt{2\pi n}(\frac{n}{e})^n < n! < \sqrt{2\pi
n}(\frac{n}{e})^ne^{\frac{1}{12}}$, (\ref{equ.proof_1.7}) can be further derived  as $
i^* - \ln(\sqrt{2 \pi i^*} e^{\frac{1}{12}}) < \frac{\lambda \pi r_{c}^2}{\beta}, i^* + 1 - \ln(\sqrt{2 \pi i^*}) > \frac{\lambda \pi r_{c}^2}{\beta} $. Since $i^*<N_f$, $\ln(\sqrt{2 \pi i^*}) < \ln(\sqrt{2 \pi N_f})$, we can obtain the range of $i^*$ as $\frac{\lambda
\pi r_{c}^2}{\beta} -1 \leq i^* \leq \frac{\lambda \pi r_{c}^2}{\beta} + \ln(\sqrt{2\pi N_f}) + 1$.

By
substituting  $v$ in (\ref{equ.proof_1.3}) into (\ref{equ.Lag_3}), we obtain
\begin{eqnarray}
\label{equ.proof_1.8}\textstyle
p^*_c(i)=
\begin{cases}
\frac{\beta}{\lambda \pi r_{c}^2 i^*}\sum_{j=1}^{i^*}\ln(\frac{j}{i})+\frac{1}{i^*}, &\textstyle i \leq i^* \\
0, &\textstyle i^*<i \leq N_f.\\
\end{cases}
\end{eqnarray}

Finally, we prove Proposition \ref{p:1}.
 
\section{Proof of Proposition \ref{t:1}}
\label{a:t1}
The cumulative distribution function (cdf) of the distance between a DR
requesting the $i$th file and the nearest helper in the $i$th \emph{helper set} can be obtained as $F_i(r)=1-e
^{-\lambda_i \pi r^2}$ \cite{SKM.PPP}. Then, the pdf of the D2D link distance can be obtained as
$f_i(r)=\frac{\text{d}F_i(r)}{\text{d}r}=2\pi r \lambda_i e^{-\lambda_i \pi r^2}$.

According to the definition, the offloading probability can be obtained as
\begin{equation}
\label{b:1.1}
\begin{aligned}\textstyle
p_1(P_t,\rho) = \sum_{i=1}^{N_f} p_r(i) \int_{0}^{r_c} f_i(r) \mathbb{P} \left[E_1(i,r) \leq \rho V_0 Q \right] dr,
\end{aligned}
\end{equation}
where $\mathbb{P}[\cdot]$ is the probability of an event.

Denoting $\Gamma_1=e^{\frac{F (P_t + \eta P_c)\ln2}{W\rho Q V_0 \eta }} - 1$, the energy constraints $E_1(i,r) \leq \rho V_0 Q$ can be rewritten as $\frac{hr^{-\alpha}}{I_{i,r}+\sigma_0^2} \geq \Gamma_1$, and \eqref{b:1.1} can be further written as
\begin{equation}\label{b:1.2}
\begin{aligned}
p_1(P_t, \rho) &\textstyle = \sum_{i=1}^{N_f} p_r(i) \int_{0}^{r_c} f_i(r) \mathbb{P} \left[\frac{hr^{-\alpha}}{I_{i,r}+\sigma_0^2} \geq \Gamma_1 \right] dr.
\end{aligned}
\end{equation}
Since $h$ follows an exponential distribution with unit mean, $\mathbb{P} \left[\frac{hr^{-\alpha}}{I_{i,r}+\sigma_0^2} \geq \Gamma_1 \right]$ can be obtained as
\begin{equation}
\label{b:1.3}
\begin{aligned}\textstyle
\mathbb{P} \left[ h \geq \Gamma_1 r^{\alpha} (\sigma_0^2 + I_{i,r}) \right] & \stackrel{(a)}{=} \mathbb{E}_{I_{i,r}} \left[ \exp\left( -r^{\alpha}\Gamma_1(\sigma_0^2 + I_{i,r}) \right)  \right]  = e^{-\Gamma_1r^{\alpha}\sigma_0^2} \mathcal{L}_{I_{i,r}}\left(r^{\alpha}\Gamma_1\right),
\end{aligned}
\end{equation}
where (a) is obtained using the same method as in \cite{PPP}, and $\mathcal{L}_{I_{i,r}}(s)$ is the Laplace transform of the random variable $I_{i,r}$. Note that if some DTs interrupt during transmission, the interference should be lower. However, we ignore this situation to make the analysis tractable to reflect the essential insights.
%

To derive $ \mathcal{L}_{I_{i,r}}\left(r^{\alpha}\Gamma_1\right)$, we need to obtain the probability that a helper  acts as a DT, called active probability.
%

As mentioned in Section \ref{s:3}, the users in the $i$th helper set follows a PPP with density $\lambda_i$. Since it is hard to directly derive the active probability, we derive its complementary probability. We first derive the  probability that a helper cached with the $i$th file does not act as a DT, denoted as $p_s(i)$. Considering that the coverage area of each helper can not exceed $\pi r_c^2$ and the probability that no DRs requesting the $i$th file are located in the coverage with area $x$ of a helper cached with the $i$th file is $e^{-\lambda p_r(i) x}$, $p_s(i)$ can be derived as
\begin{equation}
\label{b:1.6}
\begin{aligned}
\textstyle
p_s(i) &\textstyle= \frac{\int_{0}^{\pi r_c^2} e^{-\lambda p_r(i) x}g_x(x,\lambda_i) dx}{\int_{0}^{\pi r_c^2} g_x(x,\lambda_i) dx} \textstyle=\frac{\Gamma(3.5,0)-\Gamma\left(3.5,\left(3.5\lambda_i+\lambda p_r(i)\right) \pi r_c^2\right)}{\left(3.5\lambda_i+\lambda p_r(i)\right)^{3.5}} \frac{(3.5\lambda_i)^{3.5}}{\Gamma(3.5,0)-\Gamma(3.5,3.5\lambda_i \pi r_c^2)}\\
&\textstyle = \left(1+\frac{\lambda p_r(i)}{3.5\lambda_i}\right)^{-3.5} \frac{\Gamma(3.5,0)-\Gamma\left(3.5,\left(3.5\lambda_i+\lambda p_r(i)\right) \pi r_c^2\right)}{\Gamma(3.5,0)-\Gamma(3.5,3.5\lambda_i \pi r_c^2)} = \left(1+\frac{\lambda p_r(i)}{3.5\lambda_i}\right)^{-3.5}\theta_i,
\end{aligned}
\end{equation}
where $\theta_i \triangleq \frac{\Gamma(3.5,0)-\Gamma\left(3.5,\left(3.5\lambda_i+\lambda p_r(i)\right) \pi r_c^2\right)}{\Gamma(3.5,0)-\Gamma(3.5,3.5\lambda_i \pi r_c^2)}$, $\Gamma(s,x) = \int_{x}^{\infty}t^{s-1}e^{-t}dt$ is the upper incomplete Gamma function \cite{abramowitz1964}, $g_x(x,\lambda_i)$ is the pdf of the coverage area for a typical Voronoi cell in a Poisson random tessellation, which can be fitted  as $g_x(x,\lambda_i) =  \frac{3.5^{3.5}}{\Gamma(3.5,0)}\lambda_i^{3.5}x^{2.5}e^{-3.5\lambda_ix}$ \cite{ferenc2007size}.

Then, for the probabilistic caching policy, the active probability can be obtained as
\begin{equation}
\label{b:1.7}
p_a  = 1- \sum_{i=1}^{N_f} p_c(i) p_s(i).
\end{equation}

The DTs cached with the $i$th file follow a PPP distribution with density $\lambda^d_i =(1-p_s(i))p_c(i)\lambda = (1-p_s(i))\lambda_i$ \cite{PPPactive}, and the density of all DTs is $\lambda_I = \sum_{i=1}^{N_f}\lambda_i^d = \sum_{i=1}^{N_f} (1-p_s(i))p_c(i)\lambda = p_a \lambda$.

 Recall that the DR establishes a D2D link with the nearest helper cached with the desired file. Then, for the DR requesting the $i$th file, the DTs cached with other files with density $\lambda_I-\lambda^d_i$ can be closer than the desired DT. 
 
 Using Theorem 2 in \cite{PPP}, when $\alpha>2$, $ \mathcal{L}_{I_{i,r}}\left(r^{\alpha}\Gamma_1\right)$ can be derived as
\begin{equation}
\label{b:1.8}
\begin{aligned}
& \textstyle\mathcal{L}_{I_{i,r}}\left(r^{\alpha}\Gamma_1\right)  = \exp\left(-2\pi\left(\lambda_I-\lambda_i^d \right) \int_{0}^{ \infty} \frac{\Gamma_1}{\Gamma_1+(v/r)^{\alpha}} v dv -2\pi\lambda_i^d \int_{r}^{ \infty} \frac{\Gamma_1}{\Gamma_1+(v/r)^{\alpha}} v dv   \right)\\
& \textstyle = \exp\left(-2\pi\lambda_I \int_{0}^{ \infty} \frac{\Gamma_1}{\Gamma_1+(v/r)^{\alpha}} v dv +2\pi\lambda_i^d \int_{0}^{ r} \frac{\Gamma_1}{\Gamma_1+(v/r)^{\alpha}} v dv   \right)\\
& \textstyle \stackrel{}{=} \exp\left(-\pi\lambda_Ir^2 \Gamma_1^{2/\alpha} \int_{0}^{ \infty} \frac{1}{1+u^{\alpha/2}} du + \pi\lambda_i^d r^2 \Gamma_1^{2/\alpha} \int_{0}^{\Gamma_1^{-\frac{2}{\alpha}}} \frac{1}{1+u^{\alpha/2}} du\right) = \exp\left(-\pi r^2 \Gamma_1^{2/\alpha} (\lambda_I\xi_{1}-\lambda_i^d\xi_{2}) \right),
\end{aligned}
\end{equation}
where $\xi_1 =\int_{0}^{ \infty} \frac{1}{1+u^{\alpha/2}} du $ and $\xi_2 =\int_{0}^{\Gamma_1^{-\frac{2}{\alpha}}} \frac{1}{1+u^{\alpha/2}} du $. 

By substituting \eqref{b:1.3} and \eqref{b:1.8} into \eqref{b:1.2}, we prove Proposition \ref{t:1}.

\section{Proof of Proposition \ref{sp:1}}
\label{a:sp1}
Since \eqref{b:1.8} holds only when $\alpha>2$, in case of $\alpha=2$, we need to re-derive the Laplace transform $ \mathcal{L}_{I_{i,r}}\left(r^{\alpha}\Gamma_1\right)$. By ignoring the interference generated by the DTs with distance larger than $r_{\max}$ as in \cite{lichte2010expected}, we can approximate $ \mathcal{L}_{I_{i,r}}\left(r^{\alpha}\Gamma_1\right)$ as
\begin{equation}
\label{equ.s:2.0}
\begin{aligned}
 \textstyle\mathcal{L}_{I_{i,r}}\left(r^{\alpha}\Gamma_1\right)  & \textstyle \stackrel{}{\approx} \exp\left(-2\pi\left(\lambda_I-\lambda_i^d \right) \int_{0}^{ r_{\max}} \frac{\Gamma_1}{\Gamma_1+(v/r)^{2}} v dv -2\pi\lambda_i^d \int_{r}^{ r_{\max}} \frac{\Gamma_1}{\Gamma_1+(v/r)^{2}} v dv   \right)\\
& \textstyle \stackrel{(a)}{\approx} \exp\left(-\pi\lambda_Ir^2 \Gamma_1 \int_{0}^{ r_{\max}} \frac{1}{1+u} du \right) = \exp\left(-\pi r^2 \Gamma_1 \lambda_I \ln(1+r_{\max})\right),
\end{aligned}
\end{equation}
where (a) is obtained by using $\lambda_I \gg \lambda_i^d$ when the file catalog size $N_f$ is large. 

Then, the offloading probability can be obtained by substituting $\mathcal{L}_{I_{i,r}}\left(r^{\alpha}\Gamma_1\right)$ into \eqref{b:1.3} as
\begin{equation}
\label{equ.s:2.1}
\begin{aligned}\textstyle
&p_1(P_t,\rho)\textstyle \approx \sum_{i=1}^{N_f} p_r(i) \int_{0}^{r_c}  2\pi\lambda_ire^{-\pi\lambda_ir^2}e^{-\Gamma_1r^2\sigma_0^2 - \pi \lambda_I\ln(1+r_{\max}) r^2\Gamma_1 }dr  \textstyle=\sum_{i=1}^{N_f}  \frac{p_r(i) \pi\lambda_i}{\varphi_i(P_t)}(1 - e^{- \varphi_i(P_t) r_c^2}),
\end{aligned}
\end{equation}
where $\varphi_i(P_t) = \sigma_0^2\Gamma_1 + \pi \lambda_I\xi_s\Gamma_1+ \pi\lambda_i$, and $\xi_s = \ln(1+r_{\max})$. 

By denoting $g_i(P_t) =  \frac{1 - e^{-\varphi_i(P_t)r_c^2}}{\varphi_i(P_t)} $, and taking the derivative of $g_i(P_t)$ with respect to $P_t$, we can obtain
\begin{equation}
\label{equ.sp1:1}
\begin{aligned}\textstyle
g_i'\left(P_t\right) &\textstyle=  \frac{\varphi_i'\left(P_t\right)r_c^2e^{-\varphi_i\left(P_t\right)r_c^2}\varphi_i\left(P_t\right) -\varphi_i'\left(P_t\right)\left(1-e^{-\varphi_i\left(P_t\right)r_c^2}\right) }{\left(\varphi_i\left(P_t\right)\right)^2} \textstyle =  \underbrace{\frac{ \left( \kappa\left(\varphi_i\left(P_t\right)r_c^2\right)-1\right) }{\left(\varphi_i\left(P_t\right)\right)^2}}_{(\text{I})}\underbrace{\varphi_i'\left(P_t\right)}_{(\text{II})},
\end{aligned}
\end{equation}
where $\kappa(t) \triangleq (1+t)e^{-t}$, $t \geq 0$. It is not hard to show that $\kappa'(t) = -te^{-t} \leq 0$, so $\kappa(t)$ is a decreasing function of $t$ and $\kappa(0) = 1$ is the maximal value of $\kappa(t)$. Therefore, $\kappa(t) \leq 1$ and the equality holds when $t=0$. Because $\varphi_i\left(P_t\right)r_c^2 >0$, $\kappa\left(\varphi_i\left(P_t\right)r_c^2\right) < 1$ always holds. Then, part (I) in \eqref{equ.sp1:1} $\frac{ \left( \kappa\left(\varphi_i\left(P_t\right)r_c^2\right)-1\right) }{\left(\varphi_i\left(P_t\right)\right)^2} < 0$ always holds.

By changing variable $P_t \rightarrow x$ and denoting $a=\frac{F\ln2}{W\rho QV_0\eta}$, we have $\varphi_i(x) = \sigma^2\frac{(e^{a(x+\eta P_c)}-1)}{x} + \pi \lambda_I\xi_s(e^{a(x+\eta P_c)}-1)+ \pi\lambda_i$, whose first-order derivative can be obtained as
\begin{equation}
\label{equ.sp1:2}
\begin{aligned}\textstyle
\varphi_i'(x) &\textstyle=\underbrace{\frac{\sigma^2}{x^2}\left(e^{a(x+\eta P_c)}ax-e^{a(x+\eta P_c)}+1\right)}_{u_1(x)} + \underbrace{a\pi\lambda_I\xi_se^{a(x+\eta P_c)}}_{u_2(x)}.
\end{aligned}
\end{equation}

The first-order derivative of $u_1(x)$ can be derived as
\begin{equation}
\label{equ.sp1:3}
\begin{aligned}\textstyle
u_1'(x) &\textstyle=  \frac{1}{x^3}\underbrace{(a^2e^{a(x+\eta P_c)}x^2 - 2ae^{a(x+\eta P_c)}x + 2e^{a(x+\eta P_c)}-2)}_{v(x)}.
\end{aligned}
\end{equation}

It is not hard to obtain that $v'(x) = a^3e^{a(x+\eta P_c)}x^2 \geq0$. Therefore, $v(x)$ is an increasing function of $x$ and $v(0) = 0$. Because $x>0$, we know that $v(x)\geq 0$ always holds, i.e., $u_1'(x) \geq 0$, and hence $u_1(x)$ an increasing function. 

By using the same approach, we can show that $u_2(x) = a\pi\lambda_I\xi_se^{a(x+\eta P_c)}$ in \eqref{equ.sp1:2} is an increasing function. Then, $\varphi_i'(x)$ is an increasing function. Besides, when $x \to 0$, $\lim_{x \to 0} \varphi_i'(x) = \lim_{x \to 0}(a\pi\lambda_I\xi_se^{a\eta P_c} - \frac{(e^{a\eta P_c}-1)\sigma^2}{x^2}) = -\infty$. 

When $x = P_{\max}$,  $\varphi_i'(P_{\max}) = \frac{\sigma^2}{P_{\max}^2}(e^{a(P_{\max}+\eta P_c)}aP_{\max}-e^{a(P_{\max}+\eta P_c)}+1 ) + a\pi\lambda_I\xi_se^{a(P_{\max}+\eta P_c)} > \frac{\sigma^2}{P_{\max}^2}(-e^{a(P_{\max}+\eta P_c)}) +a\pi\lambda_I\xi_se^{a(P_{\max}+\eta P_c)} = e^{a(P_{\max}+\eta P_c)}(a\pi\lambda_I\xi_s - \frac{\sigma^2}{P_{\max}^2})> 0$, because $\frac{\sigma^2}{P_{\max}^2} \ll 1$ in practice.
Further considering that part (I) in \eqref{equ.sp1:1} is negative, $g_i'(P_t)$ is first greater than zero and then less than zero, and hence $g_i(P_t)$ first increases and then decreases with $P_t$. Therefore, the global optimal transmit power $P_i^*$ to maximize $g_i(P_t)$ can be obtained by solving $\varphi_i'(P_t) = 0$.
It is worthy to  note that although $\varphi_i(x)$ depends on $i$, its first-order derivative does not. Therefore, the optimal transmit power $P_t^*$ to maximize $p_1(P_t,\rho)$ in \eqref{equ.s:2.1} is the same as that to maximize $g_i(P_t)$.

Considering that today's device battery capacity is usually large, the file size is not very large (typically less than 3 GBytes), and $\frac{\rho QV_0\eta}{P_{\max}+\eta P_c^T}$ is the maximal time that a DT can transmit with $P_{\max}$, we have $a(P_{\max}+\eta P_c)= \frac{F\ln2(P_{\max}+\eta P_c)}{W\rho QV_0\eta} = F\ln2 \frac{1}{W} \frac{P_{\max}+\eta P_c}{\rho QV_0\eta} \ll 1$. By using the approximation $e^{t} \approx 1+t$, when $t \ll 1$, $ \varphi_i'(x) ={\frac{\sigma^2}{x^2}\left(e^{a(x+\eta P_c)}ax-e^{a(x+\eta P_c)}+1\right)} + {a\pi\lambda_I\xi_se^{a(x+\eta P_c)}}
\textstyle \approx {\frac{\sigma^2}{x^2}\left((a(x+\eta P_c)+1)ax-(a(x+\eta P_c)+1)+1\right)} + {a\pi\lambda_I\xi_s(a(x+\eta P_c)+1)}.$

Then, the optimal $x^*$ satisfying $\varphi_i'(x) = 0$ can be obtained by solving the cubic equation $a^2\mu x^3 + \left( a(\sigma^2 +\mu \eta P_c ) + \mu \right) ax^2 +a^2 \eta P_c\sigma^2x-a \eta P_c\sigma^2 = 0$,
where $\mu = \pi\lambda_I\xi_s$. From the equation, we can obtain the closed-form of $P_t^*$. 

This proves Proposition \ref{sp:1}.

\section{Proof of Proposition \ref{p:2}}
\label{a:p2}
Denote $\delta_1(i,r)$ as the ratio of the data conveyed via D2D links to the file size $F$, which can be obtained as $\delta_1(i,r)= \min \left(R_1(i,r)\frac{\rho V_0 Q}{F\left(\frac{1}{\eta}P_t+P_c\right)},1\right) =  \min \left( \log_2\left(1+\gamma_1(i,r)\right)\frac{W\rho V_0 Q}{F\left(\frac{1}{\eta}P_t+P_c\right)},1\right)$. From the definition, the offloading ratio can be obtained as
\begin{equation}
\label{equ.p_a1}
\begin{aligned}
& p^a_1(P_t,\rho)  \textstyle = \sum_{i=1}^{N_f} p_r(i) \int_{0}^{r_c} f_i(r)  \mathbb{E}_h\left[\delta_1(i,r)\right]  dr \\
&  \textstyle = \sum_{i=1}^{N_f} p_r(i) \int_{0}^{r_c} f_i(r)  \mathbb{P} \left[\delta_1(i,r) = 1 \right]   dr   \textstyle+ \sum_{i=1}^{N_f} p_r(i) \int_{0}^{r_c} f_i(r) \mathbb{P} \left[\delta_1(i,r) < 1 \right] \mathbb{E}_h\left[\delta_1(r)\, | \,\delta_1(i,r) < 1 \right]  dr\\
&  \textstyle \stackrel{(a)}{=} \sum_{i=1}^{N_f} p_r(i) \int_{0}^{r_c} f_i(r)  \mathbb{P} \left[ E_1(i,r) \leq \rho V_0 Q \right]   dr \\
& \textstyle+ \sum_{i=1}^{N_f} p_r(i) \int_{0}^{r_c} f_i(r) \mathbb{P} \left[\delta_1(i,r) < 1 \right] \mathbb{E}_h\left[\delta_1(i,r)\, | \,\delta_1(i,r) < 1 \right]  dr\\
& \textstyle = p_1(P_t,\rho)  + \sum_{i=1}^{N_f} p_r(i) \int_{0}^{r_c} f_i(r) \mathbb{P} \left[\delta_1(i,r) < 1 \right] \mathbb{E}_h\left[\delta_1(i,r)\, | \,\delta_1(i,r) < 1 \right]  dr,
\end{aligned}
\end{equation}
where (a) comes from the fact that $\mathbb{P} \left[\delta_1(i,r) = 1 \right] = \mathbb{P} \left[ E_1(i,r) \leq \rho V_0 Q \right] $.

From the expression of $\delta_1(i,r)$ and $R_1(i,r)$, we have
\begin{equation}
\begin{aligned}\textstyle
& \textstyle\mathbb{E}_h\left[\delta_1(i,r)\, | \,\delta_1(i,r) < 1 \right] = \mathbb{E}_h\left[\log_2\left(1+\gamma_1(i,r)\right)\frac{W\rho V_0 Q}{F\left(\frac{1}{\eta}P_t+P_c\right)} \, | \, \log_2\left(1+\gamma_1(i,r)\right)\frac{W\rho V_0 Q}{F\left(\frac{1}{\eta}P_t+P_c\right)} < 1 \right]\\
&\textstyle \stackrel{(a)}{=} \mathbb{E}_h\left[\frac{ \ln\left(1+\gamma_1(i,r)\right) }{ \ln\left(1+\Gamma_1\right) } \, | \, \frac{ \ln\left(1+\gamma_1(i,r)\right) }{ \ln\left(1+\Gamma_1\right) } < 1 \right] \textstyle= \frac{1}{\ln(1+\Gamma_1)} \mathbb{E}_h\left[\ln(1+\gamma_1(i,r))\, | \,\ln(1+\gamma_1(i,r)) < \ln(1+\Gamma_1) \right],\nonumber
\end{aligned}
\end{equation}
where (a) is obtained by substituting $\ln(1+\Gamma_1) = \frac{F (P_t + \eta P_c)\ln2}{W\rho Q V_0 \eta }$, and $\Gamma_1=e^{\frac{F (P_t + \eta P_c)\ln2}{W\rho Q V_0 \eta }} - 1$.

For a positive random variable $x$ with cdf $F(x)$ and pdf $f(x)$, we have
\begin{equation}
\label{ap1:1.2} 
\begin{aligned}\textstyle
\mathbb{E}\left[x\, | \,x < X_0\right] & \textstyle \stackrel{(a)}{=} \int_{0}^{X_0} x \frac{f(x)}{F(X_0)}dx \stackrel{}{=} X_0 - \frac{1}{F(X_0)} \int_{0}^{X_0} \mathbb{P}\left[ x<t \right] dt ,
\end{aligned}
\end{equation}
where (a) comes from the fact that the conditional pdf of $x$ given $x<X_0$ is  $\frac{f(x)}{F(X_0)}$.

Then, we have
\begin{equation}
\label{ap1:1.3} 
\begin{aligned}
&\textstyle\mathbb{E}_h\left[\ln(1+\gamma_1(i,r))\, | \,\ln(1+\gamma_1(i,r)) < \ln(1+\Gamma_1) \right] \\
&\textstyle= \ln(1+\Gamma_1) - \frac{1}{\mathbb{P}[\ln(1+\gamma_1(i,r)) < \ln(1+\Gamma_1)]} \int_{0}^{\ln(1+\Gamma_1)} \mathbb{P}[\ln(1+\gamma_1(i,r)) < t] dt\\
&\textstyle\stackrel{(a)}{=}\ln(1+\Gamma_1) - \frac{1}{\mathbb{P}[h< \Gamma_1 (\sigma_0^2 + I_{i,r}) r^{\alpha}]} \int_{0}^{\ln(1+\Gamma_1)} \mathbb{P}[h<( e^t-1)(\sigma_0^2 + I_{i,r}) r^{\alpha}] dt\\
&\textstyle\stackrel{(b)}{=} \ln(1+\Gamma_1) - \frac{1}{1-e^{-\phi_i(\Gamma_1,r)}} \int_{0}^{\ln(1+\Gamma_1)} 1-e^{-\phi_i(e^t-1,r)} dt,
\end{aligned}
\end{equation}
where (a) and (b) are respectively obtained analogous to deriving \eqref{b:1.2} and \eqref{b:1.3}, and $\phi_i(x,y) = xy^\alpha \sigma_0^2 - \pi (\lambda_I\xi_1-\lambda_i^d\xi_2) y^2 x^{2/\alpha} $.

Therefore, we have
\begin{equation}
\label{ap1:1.4}\textstyle
\mathbb{E}_h\left[\delta_1(i,r)\, | \,\delta_1(i,r) < 1 \right] = 1 - \frac{1}{1-e^{-\phi_i(\Gamma_1,r)}} \int_{0}^{\ln(1+\Gamma_1)} \frac{1-e^{-\phi_i(e^t-1,r)}}{\ln(1+\Gamma_1)} dt.
\end{equation}

On the other hand, we can obtain
\begin{equation}
\label{ap1:1.5}\textstyle
\mathbb{P} \left[\delta_1(i,r) < 1 \right] = \mathbb{P} \left[ E_1(i,r) > \rho V_0 Q \right] = 1- \mathbb{P} \left[ E_1(i,r) \leq \rho V_0 Q \right]  \stackrel{(a)}{=} 1-e^{-\phi_i(\Gamma_1,r)},
\end{equation}
where (a) is obtained according to \eqref{b:1.3}. By substituting \eqref{ap1:1.5} and \eqref{ap1:1.4} into \eqref{equ.p_a1}, we can obtain the expression of $p_1^a(P_t,\rho)$ in Proposition \ref{p:2}.

From \eqref{equ.p_a1}, we can show that $p_1(P_t,\rho) \leq p_1^a(P_t,\rho)$. Considering $\delta_1(i,r) \leq 1$, we can obtain $ p^a_1(P_t,\rho)  \textstyle = \sum_{i=1}^{N_f} p_r(i) \int_{0}^{r_c} f_i(r)  \mathbb{E}_h\left[\delta_1(i,r)\right]  dr \leq \sum_{i=1}^{N_f} p_r(i) \int_{0}^{r_c} f_i(r) dr = p_o$. Then, $p_1(P_t,\rho) \leq p_1^a(P_t,\rho) \leq p_o$. When $\phi_i(\Gamma_1,r) = 0$, according to Proposition \ref{t:1}, $p_1(P_t,\rho) = \sum_{i=1}^{N_f} p_r(i) \int_{0}^{r_c} f_i(r) e^{0} dr = \sum_{i=1}^{N_f} p_r(i) (1-e^{\pi\lambda_ir_c^2}) dr =  p_o$ and both equalities hold, where the conditions that lead to $\phi_i(\Gamma_1,r) = 0$ can be further derived as follows.

From the expression of $\Gamma_1$, we have
$\lim\limits_{\rho \rightarrow \infty} \Gamma_1 =\lim\limits_{\rho \rightarrow \infty} e^{\frac{F (P_t + \eta P_c)\ln2}{W\rho Q V_0 \eta }} - 1 = 0.$ Hence, $ \lim\limits_{\rho \rightarrow \infty}\phi_i(\Gamma_1,r) = \lim\limits_{\rho \rightarrow \infty} \Gamma_1r^{\alpha}\sigma_0^{2}+\pi (\lambda_I\xi_1-\lambda_i^d\xi_2) r^2  \Gamma_1^{2/a} = 0$.
Since $0 \leq \lambda_i^d  \leq \lambda_I$, $\lambda_I \rightarrow 0$ leads to $\lambda_i^d \rightarrow 0$.
Further considering that $\text{SNR} = P_t r^{-\alpha}/ \sigma^2 =1/r^{\alpha}\sigma_0^2$, we have $\lim\limits_{\begin{subarray}
	\text{SNR} \rightarrow \infty \\ \;\;\;\lambda_I \rightarrow 0
	\end{subarray} } \phi_i(\Gamma_1,r)  = 0.
$
Therefore, the upper bound of $p^a_1(P_t,\rho)$ can be achieved when $\rho \rightarrow \infty$, or when $\text{SNR} \rightarrow \infty$ and $\lambda_I \rightarrow 0$.
 
\section{Proof of Proposition \ref{t:2}}
\label{a:t2}
The average energy consumed at a DT for complete transmission can be obtained as
\begin{equation}
\label{e:1.a1}
\begin{aligned}\textstyle
\bar{E}_1 = \sum_{i=1}^{N_f} p'_r(i) \int_{0}^{r_c} f'_i(r) \mathbb{E}_h\left[E_1(i,r)\, | \,E_1(i,r) < \rho V_0 Q \right] \,dr,
\end{aligned}
\end{equation}
where $p'_r(i)$ is the probability that the $i$th file is requested by a satisfied DR, $f'_i(r)$ is the pdf of the D2D link distance for a DR that requests the $i$th file and is satisfied.

We can obtain $p_r'(i)$ as $p_r'(i) = \frac{p_r(i)\int_{0}^{r_c}f_i(r) \mathbb{P} \left[E_1(i,r) \leq \rho V_0 Q \right] dr}{p_1(P_t,\rho)} \stackrel{(a)}{=} \frac{p_r(i)\int_{0}^{r_c}f_i(r) e^{ -\phi_i(\Gamma_1,r) } dr} {p_1(P_t,\rho)} $,  where $p_r(i)\int_{0}^{r_c}f_i(r) \mathbb{P} \left[E_1(i,r) \leq \rho V_0 Q \right] dr$ is the probability that a DR requests the $i$th file and can be satisfied, $p_1(P_t,\rho)$ is the probability that a DR requesting any file can be satisfied, and (a) is obtained analogously to deriving \eqref{b:1.1}.  The cdf of the D2D distance for a satisfied DR that requests the $i$th file can be obtained as $F'_i(R) =  \mathbb{P} [r\leq R]= \frac{\int_{0}^{R}f_i(r) \mathbb{P} \left[E_1(i,r) \leq \rho V_0 Q \right] dr}{\int_{0}^{r_c}f_i(r) \mathbb{P} \left[E_1(i,r) \leq \rho V_0 Q \right] dr}= \frac{\int_{0}^{R}f_i(r) e^{ -\phi_i(\Gamma_1,r)  } dr}
{\int_{0}^{r_c}f_i(t) e^{ -\phi_i(\Gamma_1,r)  } dr }$, where $\int_{0}^{R}f_i(r) \mathbb{P} \left[E_1(i,r) \leq \rho V_0 Q \right] dr$ is the probability that a DR that desires the $i$th file can be satisfied with a D2D transmission distance smaller than $R$, and $\int_{0}^{r_c}f_i(r) \mathbb{P} \left[E_1(i,r) \leq \rho V_0 Q \right] dr$ is the probability that a DR desiring the $i$th file can be satisfied. Then, the pdf  $f_i'(r)$  can be obtained as $f_i'(r) = \frac{\text{d}F'_i(r)}{dr} = \frac{f_i(r) e^{ -\phi_i(\Gamma_1,r)  } } {\int_{0}^{r_c}f_i(r) e^{ -\phi_i(\Gamma_1,r)  } dr }$.

Considering that $E_1(i,r) = \frac{\ln(1+\Gamma_1)\rho V_0 Q}{\ln(1+\gamma_1(i,r))}$, we have
\begin{equation}
\label{e:1.1}\textstyle
\mathbb{E}_h\left[E_1(i,r)\, | \,E_1(i,r) < \rho V_0 Q \right]  = \ln(1+\Gamma_1)\rho V_0 Q \mathbb{E}_h\left[\frac{1}{\ln\left(1+\gamma_1\left(i,r\right)\right)}\, | \,\frac{1}{\ln\left(1+\gamma_1\left(i,r\right)\right)} < \Gamma_1'\right],
\end{equation}
where $\Gamma_1' = \frac{1}{\ln(\Gamma_1+1)}$.

Moreover, the expectation in \eqref{e:1.1} can be derived as
\begin{equation}
\label{e:1.3}
\begin{aligned}
&\textstyle\mathbb{E}_h\left[\frac{1}{\ln\left(1+\gamma_1\left(i,r\right)\right)}\, | \,\frac{1}{\ln\left(1+\gamma_1\left(i,r\right)\right)} < \Gamma_1'\right]\stackrel{(a)}{=} \Gamma_1' - \frac{1}{\mathbb{P}\left[ \frac{1}{\ln\left(1+\gamma_1\left(i,r\right)\right)} <\Gamma_1' \right]} \int_{0}^{\Gamma_1'} \mathbb{P}\left[ \frac{1}{\ln\left(1+\gamma_1\left(i,r\right)\right)} <t \right] dt \\
&\textstyle\stackrel{(b)}{=} \Gamma_1' - \frac{1}{\mathbb{P}\left[  h> \Gamma_1' (\sigma_0^2 + I_{r,i}) r^{\alpha}    \right]} \int_{0}^{\Gamma_1'} \mathbb{P}\left[  h> (e^{\frac{1}{t}} -1)(\sigma_0^2 + I_{r,i}) r^{\alpha}  \right] dt\\
&\textstyle\stackrel{(c)}{=}\Gamma_1' - \frac{ 1   }{   e^{ -\Gamma_1'\sigma_0^2  r^{\alpha} -\pi (\lambda_I\xi_1-\lambda_i^d\xi_2) r^2 (\Gamma_1')^{\frac{2}{\alpha}}  }  } \int_{0}^{\Gamma_1'} e^{ -(e^{\frac{1}{t}} -1)\sigma_0^2  r^{\alpha} -\pi (\lambda_I\xi_1-\lambda_i^d\xi_2) r^2 (e^{\frac{1}{t}} -1)^{\frac{2}{\alpha}}  } dt\\
&\textstyle\stackrel{(d)}{=}\Gamma_1' - \frac{ 1   }{   e^{ - \phi_i(\Gamma_1',r)  }  } \int_{0}^{\Gamma_1'} e^{ \phi_i(e^{\frac{1}{t}} -1,r)   } dt,
\end{aligned}
\end{equation}
where (a) is obtained according to \eqref{ap1:1.2},  (b) is obtained by substituting the definition of $\gamma_1(i,r)$ in \eqref{equ.SINR_1}, (c) is because $h$ follows an exponential distribution with unit mean, and (d) is because $\phi_i(x,y) = xy^\alpha \sigma_0^2 - \pi (\lambda_I\xi_1-\lambda_i^d\xi_2) y^2 x^{2/\alpha} $.

By substituting $p_r'(i)$, $f_i'(r)$ and \eqref{e:1.3} into  \eqref{e:1.1} and \eqref{e:1.a1} and after some further manipulations,  Proposition \ref{t:2} follows.
 
\section{Proof of Proposition \ref{pt:4}}
\label{a:pt4}
By denoting $A=r^{\alpha}\sigma^2$, $a = \frac{F\ln2}{W\rho Q V_0 \eta }$ and $g(P_t) = \frac{A\Gamma_2}{P_t}$,  the offloading probability can be expressed as
\begin{equation}
\label{g:1.1s}\textstyle
p_2(P_t,\rho) = \sum_{i=1}^{N_f}  p_r(i) \int_{0}^{r_c} f_i(r)  e^{-g(P_t)} dr,
\end{equation}
where  $\Gamma_2 = e^{a(x+\eta P^T_c)}-1$.

To simplify the notation, we change the variable  $P_t \rightarrow x$. We can obtain the first-order derivative of $g(x)$ as $ g'(x) = \frac{A}{x^2}{(\frac{\text{d}\Gamma_2}{\text{d}x}x-\Gamma_2)}.$
Then, the first-order derivative of $p_2(x,\rho)$ can be obtained as
\begin{equation}
\label{g:1.2s}\textstyle
p'_2(x,\rho) = -\sum_{i=1}^{N_f}  p_r(i) \int_{0}^{r_c} f_i(r)g'(x)e^{-g(x)} dr = -g_1(x)\sum_{i=1}^{N_f}  p_r(i) \int_{0}^{r_c} f_i(r)\frac{A}{x^2}e^{-g(x)} dr,
\end{equation}
where $g_1(x) =\frac{\text{d}\Gamma_2}{\text{d}x}x-\Gamma_2$. The first-order derivative of $g_1(x)$ with respect to $x$ can be obtained as $g_1'(x) = \frac{\text{d}^2\Gamma_2}{\text{d}^2x}x = a^2e^{a(x+\eta P^T_c)}x \geq 0$.

Therefore, $g_1(x)$ is an increasing function, where $g_1(0) = 1- e^{a\eta P^T_c} < 0$ and $g_1(+\infty) \rightarrow +\infty$. Then, from \eqref{g:1.2s}, we can see that $p_2(P_t,\rho)$ first increases and then decreases, and the optimal $x$ to maximize $p_2(x,\rho)$ can be obtained by solving $g_1(x)=0$. Again, considering that $a(P_{\max}+\eta P_c^T) \ll 1$ and using the approximation $e^{t} \approx 1+t$ that is accurate when $t \ll 1$, $g_1(x)$ can be derived from \eqref{g:1.2s} as
\begin{equation}
\label{g:1.4}\textstyle
g_1(x) = e^{a(x+\eta P^T_c)}(ax-1)+1 \approx (a(x+\eta P^T_c)+1)(ax-1)+1 = a^2(x^2+\eta P^T_c (x-\frac{1}{a})).
\end{equation}
Then, from $g_1(x)=0$ and $0<x\leq P_{\max}$, the optimal transmit power can be obtained as
\begin{eqnarray}
\label{g:1.5}\textstyle
P_t^*=
\begin{cases}\textstyle
P_{\max}, & \textstyle P_{\max} < \eta P_c^T \left(\sqrt{\frac{1}{a\eta P_c^T }+\frac{1}{4}}-\frac{1}{2}\right) \\
\eta P_c^T \left(\sqrt{\frac{1}{a\eta P_c^T }+\frac{1}{4}}-\frac{1}{2}\right), & \textstyle\text{otherwise} \\
\end{cases}.
\end{eqnarray}

This proves  Proposition \ref{pt:4}.

\bibliographystyle{IEEEtran}
\bibliography{CBQ_J}
\end{document}